%% file: Multibrane.tex
\documentclass[aps,eqsecnum,nofootinbib,prd,noshowpacs]{revtex4}
\usepackage{amsmath}
\usepackage{amssymb}
\usepackage{amsfonts}
\usepackage{tensor}
\usepackage{graphicx}
\usepackage{appendix}
\usepackage[pdftex,
  pdftitle={A four-dimensional description of five-dimensional N-brane models},
  pdfauthor={Jolyon Bloomfield, Eanna E Flanagan},
  bookmarks,bookmarksopen=false,
  pdfstartview={FitH}
]{hyperref}

\numberwithin{equation}{section}
\newcommand{\n}[2][{}]{\ensuremath{\tensor*[^{n}]{#2}{#1}}}
\newcommand{\np}[2][{}]{\ensuremath{\tensor*[^{n+1}]{#2}{#1}}}
\newcommand{\nm}[2][{}]{\ensuremath{\tensor*[^{n-1}]{#2}{#1}}}
\newcommand{\lrn}[1]{\ensuremath{\tensor*{#1}{_n}}}

\newcommand{\lrnp}[1]{\ensuremath{\tensor*{#1}{_{n+1}}}}

\newcommand{\kapfs}{\ensuremath{\tensor*{\kappa}{_5^2}}}
\newcommand{\bn}{{\ensuremath{{{\cal{B}}_{n}}}}}
\newcommand{\rn}{{\ensuremath{{{\cal{R}}_{n}}}}}
\newcommand{\E}{\mu}

\begin{document}

\title{A four-dimensional description of five-dimensional N-brane models}

\author{Jolyon K. Bloomfield}
\email{jkb84@cornell.edu}
\affiliation{Physics Department, Cornell University, Ithaca, New York 14853, USA}

\author{\'Eanna \'E. Flanagan}
\email{flanagan@astro.cornell.edu}
\affiliation{Center for Radiophysics and Space Research, Cornell University, Ithaca, New York 14853, USA}
\affiliation{Newman Laboratory for Elementary Particle Physics, Cornell University, Ithaca, New York 14853, USA}

\date{April 14th, 2010}

\begin{abstract}
We propose a new method for obtaining the four-dimensional effective gravitational theory for five-dimensional braneworld models with arbitrary numbers of branes in a low energy regime, based on a two-lengthscale expansion. The method is straightforward and computationally efficient, and is applicable to both compactified and uncompactified models. Particular emphasis is placed on the behavior of the radion modes of the model, while the massive effective fields are automatically truncated. Generally, the radion modes are found to form a (N-1)-dimensional nonlinear sigma model. We illustrate the method for a N-brane model in an uncompactified extra dimension.
\end{abstract}

\pacs{}

\maketitle


\renewcommand\thesection{\arabic{section}}
\renewcommand\thesubsection{\Alph{subsection}}
\renewcommand\thesubsubsection{\Roman{subsubsection}}

\input{intro}

\input{model}

\input{lengthscale}

\input{0thorder}

\input{wherenow}

\input{2ndorder}

\input{analysis}

\input{lowenergy}

\bibliography{braneworldarxiv}
\bibliographystyle{utphys}

\appendixpage
\appendix
\input{exact}

\end{document}

%% file: intro.tex
\section{Introduction and Summary}
The past ten years has seen a flurry of activity  related to extra-dimensional models. Motivated largely by string theory and M-theory, extra-dimensional models have displayed the potential to provide natural solutions to issues such as the hierarchy and cosmological constant problems, as well as providing interesting models for dark matter. Braneworld models, motivated by the works of Arkani-Hamed et al. \cite{ADD1998} and Randall and Sundrum \cite{Randall1999, Randall1999a}, have become a very active field of research, with many papers investigating extensions to the basic ideas (see, eg, \cite{Brax2003, Rubakov2001, Maartens2004} and citations therein).

Several different approximation and computational methods have been used to extract physical predictions from extra-dimensional models. In particular, many models have an effective four-dimensional regime at low energies, where the radius of curvature of spacetime measured by four-dimensional observers is much larger than a certain microphysical lengthscale. We review some of the computational methods that have been used to obtain a four-dimensional description of five-dimensional braneworld models, in order to place our results in context.

One method is to linearize the higher dimensional equations of motion about simple background solutions, then specialize to the long-lengthscale limit in order to obtain the linearized four-dimensional effective theory (which roughly corresponds to discarding the Kaluza-Klein modes). This method was used by Garriga and Tanaka \cite{Garriga2000} in their analysis of the RS-I model \cite{Randall1999}, who showed that linearized Einstein gravity is recovered on one of the branes in a particular regime. Further analyses to quadratic order and analyses on other backgrounds have also been performed; see, for example, Refs. \cite{Kudoh2001, Kudoh2002, Carena2005, Callin2005}. Linearized analyses have many advantages: they are quick and simple, and serve to identify all of the dynamical degrees of freedom in the theory, particularly the Kaluza-Klein modes. However, the linearized method is inherently limited and cannot describe strong field phenomena such as cosmology and black holes.

A second method is to project the five-dimensional equations of motion onto a brane; see, for example, Ref. \cite{Shiromizu2000}. This ``covariant curvature'' formalism fully incorporates the nonlinearities of the theory. However, the projected description includes nonlocal terms, and the truncation to a low energy effective theory is nontrivial, except in cases with high degrees of symmetry.

In order to overcome some of these shortcomings, Kanno and Soda \cite{Kanno2002a, Kanno2002, Soda2003} suggested a perturbation expansion of the covariant curvature formalism known as the ``gradient expansion method'', which involves expanding the theory in powers of the ratio between a microphysical scale and the four-dimensional curvature lengthscale. This approach allows a low-energy description of the model to be found, while retaining the nonlinearities of the theory. This method has been particularly successful in investigating the cosmology of braneworld models \cite{Kanno2002a, Soda2003, Zen2005, Arroja2008}, and has the benefit of providing an explicit calculation of the five-dimensional metric, but is algebraically complex and requires assumptions on the form of the metric.

An alternative approach to obtaining a four-dimensional effective action, discussed by Wiseman \cite{Wiseman2002}, focusses on the radion mode of the RS-I model. Treating the radion mode as a deflection of the branes, the approach uses a derivative expansion to calculate its nonlinear behavior. Although this method nicely captures the nonlinearities of the theory, it is highly nontrivial, and ``guesses'' the four-dimensional effective action, based on the first order equations of motion the method finds.

A final method involves making an ansatz for the form of the five-dimensional metric in terms of four-dimensional fields, and integrating over the fifth dimension to obtain a four-dimensional action. Examples of this method in the literature include Refs. \cite{Arroja2008, Chiba2000, Goldberger2000, Cotta-Ramusino2004, Kanno2005a}. The benefits of this method are the automatic truncation of the massive Kaluza-Klein modes, and the computational efficiency in dealing strictly at the level of the action. The main drawback is that the five-dimensional metric ansatz must be found (or guessed) using another method.

Four-dimensional effective descriptions typically contain moduli fields (radion modes) which describe the distances between branes. Often, such modes appear as massless scalar fields which couple to gravity in a Brans-Dicke like manner (see, eg, Ref. \cite{Chiba2000}). This occurs in the RS-I model of two branes in a compactified bulk with orbifold symmetry, for example. In this model, the radion mode must be stabilized by some mechanism (for example, by using a bulk scalar field as in the Goldberger-Wise mechanism \cite{Goldberger1999a}), or else the theory is ruled out for observers on the TeV brane (see, eg, Refs. \cite{Garriga2000, Bean2008a}). In theories including multiple branes, one expects several radion modes which may have nontrivial couplings to one another and to the four-dimensional metric at the nonlinear level.

One common extension of the RS models is to consider models with more than one or two branes. A variety of papers have considered three-brane models, usually on an orbifold (see, eg, \cite{Zen2005, Cotta-Ramusino2004, Kogan2000, Kogan2001, Kogan2002}). Some special cases have been considered for arbitrary $N$-brane models, mostly to investigate their cosmological properties \cite{Kogan2001, Flanagan2001a}. A few papers comment that their methods should extend to arbitrary $N$-brane situations (eg, \cite{Damour2002}), but little analysis has actually been performed in this regard.

In this paper, we present a new method to obtain a four-dimensional effective theory from an $N$-brane model in five dimensions. We assume that matter is confined to branes with the only bulk field being gravity, and we do not invoke mechanisms to stabilize the radion modes. The method utilizes a two-lengthscale expansion to find solutions to the five-dimensional equations of motion in a low energy regime. We do not require assumptions about the form of the metric, or the existence of Gaussian normal coordinates. The method is computationally efficient, and does not require the explicit use of the five-dimensional Einstein equations or the Israel junction conditions. Instead, one always works at the level of the action. Furthermore, our method is very general and can be applied to various models. The method has similarities to the gradient expansion method (see especially \cite{Kanno2005a}), but is computationally much simpler, and can deal with multiple branes in a straightforward manner. A particular strength of the method is that it performs a rigorous treatment of all radion modes, and automatically truncates massive modes. We present a brief example of the method for the case of the RS-I model \cite{Randall1999}, before illustrating the method in detail for the case of $N$ four-dimensional branes in an uncompactified extra dimension, deriving the four-dimensional effective action for a general configuration.

%% file: model.tex
\subsection{Applicable Models\label{SecModel}}
In this section, we introduce the parameters, metrics, and coordinate systems used to define the braneworld models which we apply our method to. The most basic model assumes that the extra dimension is infinite and not compactified, but the generalization to circularly compactified and orbifolded systems is straightforward, and is described in Section \ref{secRSI}.

We consider a system of $N$ four-dimensional branes in a five-dimensional universe with one temporal dimension, with coordinates $x^\Gamma = (x^0, \ldots, x^4)$. We denote the bulk metric by $\tensor{g}{_{\Gamma \Sigma}}(x^\Theta)$, and the associated five-dimensional Ricci scalar by $R^{(5)}$. For simplicity, we assume that there are no physical singularities in the spacetime.

The $N$ branes are labeled by an index $n = 0, 1, \ldots, N-1$, so that adjacent branes are labeled by successive values of $n$. We assume that the branes are non-intersecting. Denote the $n^{\mathrm{th}}$ brane by \bn. On \bn, we introduce a coordinate system $\tensor*{w}{_n^a} = (\tensor*{w}{_n^0}, \ldots, \tensor*{w}{_n^3})$. The location of the branes in the five-dimensional spacetime is described by $N$ embedding functions $\tensor[^n]{x}{^\Gamma}(\tensor*{w}{_n^a})$. From these embedding functions, we can calculate the induced metric \n[_{ab}]{h} on \bn,
\begin{align}
\n[_{ab}]{h}(\tensor*{w}{_n^c}) = \left.\frac{\partial \, \tensor[^n]{x}{^\Gamma}}{\partial \tensor*{w}{_n^a}} \frac{\partial \, \tensor[^n]{x}{^\Sigma}}{\partial \tensor*{w}{_n^b}} \tensor{g}{_{\Gamma \Sigma}} \left[\tensor[^n]{x}{^\Theta} \right]\right|_{\tensor*{w}{_n^c}}. \label{Embedding}
\end{align}
We associate a non-zero brane tension $\sigma_n$ with each brane \bn, and we also take there to be matter fields $\n{\phi}(\tensor*{w}{_n^a})$ which live on \bn, with their own matter action $\n[_m]{S}{}[\n[_{ab}]{h}, \n{\phi}]$.

In between each brane there exists a bulk region of spacetime, which we will denote ${\cal{R}}_0, \ldots , {\cal{R}}_N$, with \rn\ lying between branes $n-1$ and $n$. The first (last) bulk region will describe the region between the first (last) brane and spatial infinity in the bulk. In each bulk region \rn\ we allow for a bulk cosmological constant \lrn{\Lambda} (see Ref. \cite{Flanagan2001a} for a possible microphysical origin for such piecewise constant cosmological constants).

Finally, the action for the model is
\begin{align}
S \left[ \tensor{g}{_{\Gamma \Sigma}}, \tensor[^n]{x}{^\Gamma}, \n\phi \right] ={}& \int d^5 x \sqrt{-g} \left(\frac{R^{(5)}}{2 \kapfs} - \Lambda(x^\Gamma) \right) - \sum_{n = 0}^{N-1} \lrn{\sigma} \int_{\bn} d^4 \lrn{w} \sqrt{-\n{h}} + \sum_{n = 0}^{N-1} \n[_m]{S}[\n[_{ab}]{h}, \n{\phi}], \label{generalaction}
\end{align}
where \kapfs\ is the five-dimensional Newton's constant, and $\Lambda(x^\Gamma)$ takes the value \lrn{\Lambda} in \rn.

\subsection{Overview of the Method and Results\label{SecResults}}
Our method works in five steps.

\begin{figure}[t]
    \centering
        \includegraphics[width = 0.7\textwidth]{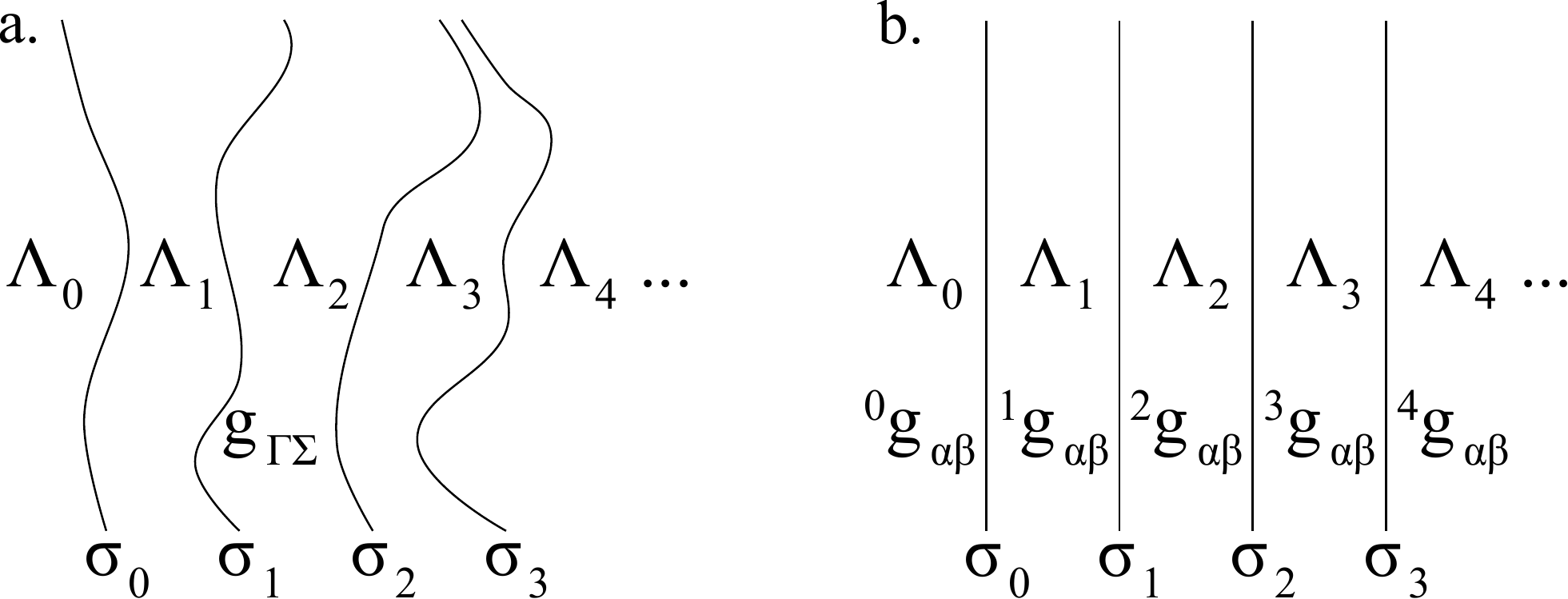}
    \caption{An illustration of the model a) before and b) after gauge fixing. The bulk cosmological constants, brane tensions, and metrics are labeled.} \label{FigSetup}
\end{figure}

\textbf{Step 1: Gauge specialize.} From the general action (Eq. \eqref{generalaction} in the model we discuss here), we perform a gauge transformation to specialize the metric to the straight gauge \cite{Carena2005}, illustrated in Figure \ref{FigSetup}.

\textbf{Step 2: Separate lengthscales in the action.} There are two characteristic lengthscales in the model. The first, which we call the microphysical lengthscale, is the lengthscale associated with the bulk cosmological constants, which is typically assumed to be on the order of the micron scale or smaller. The second lengthscale is the four-dimensional radius of curvature felt on the branes. When the ratio of the microphysical lengthscale to the four-dimensional radius of curvature is small (the low energy limit), the dynamics of the extra dimension effectively decouples from the four-dimensional dynamics, leading to a four-dimensional effective theory. We introduce a small parameter to tune this ratio, and use this parameter to perform a two lengthscale expansion of the action.

\textbf{Step 3: Solve equations of motion.} The equations of motion at zeroth order in this small parameter are calculated and explicitly solved. As expected in this type of model, all of the bulk cosmological constants must be negative, and at this order, the brane tensions are required to be tuned to a specific value\footnote{We consider small deviations from this value in Section \ref{BraneTensions}.} \cite{Randall1999} in order to avoid an effective cosmological constant on the branes. The solution to the zeroth order equations of motion provides a background metric solution, which is perturbed at the next order in our small parameter (the metric is an exact solution if the four-dimensional space is flat).

\textbf{Step 4: Integrate five-dimensional dynamics.} The five-dimensional dynamics of the theory are integrated out by substituting the metric ansatz into the action, and integrating over the extra dimension. The action to zeroth order in the small parameter is minimized by the solution, leaving only the four-dimensional terms in the action.

\textbf{Step 5: Redefine fields.} The final step is to redefine fields to cast the four-dimensional effective action in the form of a four-dimensional multiscalar-tensor theory in a nonlinear sigma model. In the Einstein conformal frame, the general form of the four-dimensional effective action is given by
\begin{align}
 S[g_{ab}, \Phi^A, \n\phi] ={}& \int d^4 x \sqrt{-g} \left\{ \frac{1}{2 \kappa_4^2} R [g_{ab}] - \frac{1}{2} \gamma_{AB}(\Phi^C) \nabla_a \Phi^A \nabla_b \Phi^B g^{ab} \right\} + \sum_{n=0}^{N-1} \n[_m]S \left[ e^{2 \alpha_n \left(\Phi^C\right)} g_{ab}, \n\phi \right] \label{generalEinsteinframe}
\end{align}
where $\Phi^A$, $1 \le A \le N-1$, are massless scalar fields (radion modes) which encode the interbrane distances. Also, $\kappa_4^2 \ (= 8 \pi G_N)$ is the effective four-dimensional Newton's constant, which is a function of $\kappa_5^2$ and the bulk cosmological constants. Finally, $\gamma_{AB}(\Phi^C)$ is the field space metric of the nonlinear sigma model, and $\alpha_n (\Phi^C)$ are the brane coupling functions. The functional form of both of these depends on the specifics of the model.

One of the features of the method used here is that five-dimensional gravitational perturbations which give rise to massive four-dimensional fields are automatically truncated. The mass scales of these fields are typically of order $\hbar/{\cal L}$, where ${\cal L}$ is the microphysical lengthscale of the theory. However, Damour and Kogan \cite{Kogan2000, Kogan2001, Damour2002} have shown that it is possible to have graviton Kaluza-Klein modes where masses are of order $\hbar/{\cal L} \exp (-l/{\cal L})$, where $l$ is an inter-brane separation. Because of the exponential factor, these second graviton modes can be ultralight and observationally relevant. Although the models we consider are likely to contain such ultralight graviton modes, our method excludes their possible contributions to a four-dimensional effective theory.

Our method has similarities to the gradient expansion method of Kanno and Soda \cite{Kanno2002a, Kanno2002, Soda2003}. Our small expansion parameter coincides with theirs, and the zeroth order solutions from both methods agree in cases where both methods are applicable. However, beyond this point, the methods diverge. Our method Taylor expands the action, but not the metric as in the covariant curvature formalism. Although higher order corrections to the metric do exist, they are intrinsically five-dimensional interactions which are unnecessary for the construction of a four-dimensional effective theory, and their contributions to the effective theory are exponentially suppressed within the low energy regime. Furthermore, our method arrives at a four-dimensional effective action, rather than working only at the level of the equations of motion. This provides for computational efficiency and a more intuitive understanding of the final result.

This paper is structured as follows. We begin by briefly illustrating our method in the context of the RS-I model in Section \ref{secRSI}. We then move on to detailing our method in the context of an uncompactified $N$-brane model with an infinite extra dimension. In Section \ref{SecCoordinates}, we specialize our coordinate systems to a convenient gauge, and in Section \ref{SecLengthscale} we discuss our two-lengthscale expansion in detail. In sections \ref{SecLowest} and \ref{SecSecond}, we solve the high energy dynamics to obtain the background metric ansatz, and derive the four-dimensional effective action of the model. We then specialize to one-brane and two-brane cases, before deriving the general $N$-brane action (Section \ref{secanalysis}). Finally, we discuss the nature of the approximations employed, before summarizing and concluding in Section \ref{endsection}.

\section{Application of the Method to the Randall Sundrum Model\label{secRSI}}
To briefly illustrate an application of our method, we apply it to the well known case of the Randall Sundrum (RS-I) model with a general background. The derivation of results in this section follows the details on the uncompactified model treated in the remainder of this paper closely.

Many papers have used a metric ansatz for the RS-I model (e.g. \cite{Randall1999, Chiba2000}), guessing at the form of the five-dimensional metric, and using this to compute the four-dimensional effective action. Such metrics are typically of the form
\begin{align}
ds^2 = e^{\chi (x^c, y)} \gamma_{ab}(x^c) dx^a dx^b + \left(\frac{\chi_{,y}(x^c, y)}{2k}\right)^2 dy^2
\end{align}
where $k = \sqrt{-\kapfs \Lambda/6}$. Rather than guessing at the form of the five-dimensional metric, our method derives a five-dimensional metric solution, from which the four-dimensional action is calculated.

The RS-I model contains two branes on a circular orbifold. We consider the circle of circumference $2L$, with the branes at $y=0$ and $y=L$, with $-L < y < L$. We will let the $y=0$ brane be the Planck brane, and the $y=L$ brane be the TeV brane. Points $y$ and $-y$ are identified. To write this in the language of regions used above, we treat the regions $-L < y < 0$ and $0 < y < L$ as two distinct regions, but identify fields by $\phi(-y) = \phi(y)$.

We now follow the computational steps outlined in Section \ref{SecResults}.

\textbf{Step 1.} Write the action in the straight gauge \cite{Carena2005}. In this gauge, the general metric is given by
\begin{align}
ds^2 = e^{\chi (x^c, y)} \gamma_{ab}(x^c, y) dx^a dx^b + \Phi^2(x^c, y) dy^2
\end{align}
where $\det \gamma = -1$, and we take $\Phi$ to be positive. The action is given by
\begin{align}
S ={}& \int d^4 x \left( \int_{0^+}^{L^-}dy + \int_{-L^+}^{0^-}dy \right) \sqrt{-g} \left(\frac{R^{(5)}}{2 \kapfs} - \Lambda \right) - \sigma_0 \int_{{\cal{B}}_0} d^4 x \sqrt{-\tensor[^0]{h}{}} - \sigma_L \int_{{\cal{B}}_L} d^4 x \sqrt{-\tensor[^L]{h}{}} \nonumber
\\
& + \frac{1}{\kapfs} \int_{{\cal{B}}_0} d^4 x \sqrt{-\tensor[^0]{h}{}} \left(\tensor[^0]{K}{^+} + \tensor[^0]{K}{^-}\right) + \frac{1}{\kapfs} \int_{{\cal{B}}_L} d^4 x \sqrt{-\tensor[^L]{h}{}} \left(\tensor[^L]{K}{^+} + \tensor[^L]{K}{^-} \right) + \tensor[^0]S{_m} \left[ \tensor[^0]h{_{ab}}, \tensor[^0]\phi{} \right] + \tensor[^L]S{_m} \left[ \tensor[^L]h{_{ab}}, \tensor[^L]\phi{} \right]. \label{Rsaction}
\end{align}
The indices $0$ and $L$ refer to the Planck and TeV branes, respectively. $h_{ab}$ is the four-dimensional induced metric on a brane, and $\sigma$ is the brane tension. $K^+$ and $K^-$ are the extrinsic curvature tensors on either side of the branes, and $S_m$ is the matter action on each brane.

\textbf{Step 2.} Now, expand the action \eqref{Rsaction} to lowest order in the two-lengthscale expansion detailed in Section \ref{SecLengthscale}. The action to lowest order in this model is given by
\begin{align}
S ={} & \int d^4 x \left( \int_{0^+}^{L^-}dy + \int_{-L^+}^{0^-}dy \right) \sqrt{-\gamma} \frac{e^{2\chi}}{2 \kapfs \Phi} \left( - \frac{1}{4} \gamma^{ab} \gamma_{bc,y} \gamma^{cd} \gamma_{da,y} - 5 \left(\chi_{,y}\right)^2 - 4 \chi_{,yy} + 4 \frac{\Phi_{,y}}{\Phi} \chi_{,y} - 2 \kapfs \Phi^2 \Lambda \right) \nonumber
\\
&+ \int d^4 x \left( \int_{0^+}^{L^-}dy + \int_{-L^+}^{0^-}dy \right) \lambda (x^a, y) \left(\sqrt{-\gamma} - 1 \right) + \int_{{\cal{B}}_0} d^4 x \sqrt{-\gamma} e^{2 \chi(0)} \left[ \frac{2}{\kapfs} \left( \left. \frac{\chi_{,y}}{\Phi} \right|_{y=0^-} - \left. \frac{\chi_{,y}}{\Phi} \right|_{y=0^+}\right) - \sigma_0 \right] \nonumber
\\
&
+ \int_{{\cal{B}}_L} d^4 x \sqrt{-\gamma} e^{2 \chi(L)} \left[ \frac{2}{\kapfs} \left( \left. \frac{\chi_{,y}}{\Phi} \right|_{y=L^-} - \left. \frac{\chi_{,y}}{\Phi} \right|_{y=-L^+}\right) - \sigma_L \right]
\label{rslowaction}
\end{align}
Here, $\chi(0)$ denotes $\chi(x^a, 0)$, and similarly for $\chi(L)$. The second line in this action includes a Lagrange multiplier ($\lambda$) to enforce the condition $\det \gamma = -1$.

\textbf{Step 3.} Varying the action \eqref{rslowaction} with respect to the three fields $\chi$, $\gamma$ and $\Phi$, the following equations of motion are obtained.
\begin{align}
0 ={}& \frac{1}{4} \gamma^{ab} \gamma_{bc, y} \gamma^{cd} \gamma_{da, y} - 3 \chi_{, y}^2 - 2 \kapfs \Phi^2 \Lambda
\\
\gamma_{ad, yy} ={}& \gamma_{ab,y} \gamma^{bc} \gamma_{cd, y} - \gamma_{ad, y} \left( 2 \chi_{,y} - \frac{\Phi_{,y}}{\Phi}\right) \label{rssecondeq}
\\
0 ={}& \frac{1}{12} \gamma^{ab} \gamma_{bc, y} \gamma^{cd} \gamma_{da, y} + \chi_{, y}^2 + \chi_{, yy} - \frac{\Phi_{,y}}{\Phi} \chi_{,y} + \frac{2}{3} \kapfs \Phi^2 \Lambda
\end{align}
The following boundary conditions at the branes are also obtained.
\begin{align}
\gamma_{ab,y}(y=0, L) ={}& 0 \label{rsfirstboundary}
\\
\chi_{,y}(y=0^+) ={}& - \frac{1}{3} \kapfs \sigma_0 \Phi \label{rssecondboundary}
\\
\chi_{,y}(y=L^-) ={}& \frac{1}{3} \kapfs \sigma_L \Phi \label{rsthirdboundary}
\end{align}

We now solve the equations of motion. The solution to \eqref{rssecondeq} is given by (in matrix notation)
\begin{align}
\boldsymbol{\gamma} (x^a, y) ={}& \mathbf{A}(x^a) \exp\left( \mathbf{B}(x^a) \int^y_{0} \Phi(x^a, y^\prime) e^{-2\chi(x^a, y^\prime)} dy^\prime \right)
\end{align}
where $\mathbf{A}$ and $\mathbf{B}$ are arbitrary $4 \times 4$ real matrix functions of $x^a$, subject to the constraint that $\boldsymbol{\gamma}$ is a metric. This can be combined with \eqref{rsfirstboundary} to yield $\mathbf{B} = \mathbf{0}$, and so $\gamma$ is a function of $x^a$ only. The only remaining equation of motion is then $\chi_{, y}^2 = - 2 \kapfs \Phi^2 \Lambda /3$. Defining $k = \sqrt{-\kapfs \Lambda/6}$, this gives $\chi_{,y} = \pm 2 k \Phi$. Choose the negative solution, so that the brane at $y=0$ corresponds to the Planck brane. The other boundary conditions \eqref{rssecondboundary} and \eqref{rsthirdboundary} yield
\begin{align}
\sigma_0 = \frac{6k}{\kapfs} \ \ \ \ \mathrm{and} \ \ \ \ \sigma_L = - \frac{6k}{\kapfs}
\end{align}
which are the well known brane tuning conditions. Combining these solutions, the metric solution is then
\begin{align}
ds^2 = e^{\chi (x^c, y)} \gamma_{ab}(x^c) dx^a dx^b + \left(-\frac{\chi_{,y}(x^c, y)}{2k}\right)^2 dy^2.
\end{align}

\textbf{Step 4.} We now have the zeroth-order metric solution, which has solved the five-dimensional dynamics. The next step is to use this metric in the original action and integrate over the fifth dimension (c.f. \cite{Chiba2000}). The zeroth order part of the action integrates to exactly zero, while the remainder of the action (the original second order terms) yields the following four-dimensional effective action.
\begin{align}
S ={}& \int d^4 x \frac{\sqrt{-\gamma}}{2k \kapfs} \left[ \left(1 - e^{\chi(L)}\right) R^{(4)} - \frac{3}{2} e^{\chi(L)} (\nabla^a \chi(L))(\nabla_a \chi(L)) \right] + \tensor[^0]S{_m} \left[ \gamma_{ab}, \tensor[^0]\phi{} \right] + \tensor[^L]S{_m} \left[ e^{\chi(L)} \gamma_{ab}, \tensor[^L]\phi{} \right]
\end{align}
The constraint $\det \gamma = -1$ has been relaxed, instead choosing $\chi(0) = 0$.

\textbf{Step 5.} Transforming to the Einstein frame, let $g_{ab} = \left(1 - e^{\chi(L)}\right) \gamma_{ab}$, and define $\exp(\chi(x^a, L)/2) = \tanh \left(\kappa_4 \varphi(x^a)/\sqrt{6}\right)$. Let $\kappa_4^2 = k \kapfs$ be the four-dimensional gravitational scale. The action in the Einstein frame is then given by
\begin{align}
S ={}& \int d^4 x \sqrt{-g} \left[ \frac{R^{(4)}}{2 \kappa_4^2} - \frac{1}{2} (\nabla^a \varphi)(\nabla_a \varphi) \right] + \tensor[^0]S{_m} \left[ \cosh^2 \left( \frac{\kappa_4 \varphi}{\sqrt{6}} \right) g_{ab}, \tensor[^0]\phi{} \right] + \tensor[^L]S{_m} \left[ \sinh^2 \left( \frac{\kappa_4 \varphi}{\sqrt{6}} \right) g_{ab}, \tensor[^L]\phi{} \right]. \label{RSIendaction}
\end{align}
This action corresponds to the four-dimensional effective action arrived at by other means, such as the covariant curvature formalism \cite{Kanno2002, Chiba2000}.\\

For the rest of this paper, we confine our discussions to uncompactified $N$-brane models.

\section{The Five-Dimensional Action in a Convenient Gauge\label{SecCoordinates}}
In this section, we take the action \eqref{generalaction} and make coordinate choices to simplify the expression. We also separate out contributions due to discontinuities in the connection across branes. We specialize the coordinate system to that of the straight gauge \cite{Carena2005}, and give the action corresponding to Eq. \eqref{generalaction} in this gauge. Again, while the details presented here are specific to an uncompactified extra dimension, they generalize straightforwardly to the other situations described previously.

In general, the five-dimensional Ricci scalar can have distributional components at the branes, as the metric will have a discontinuous first derivative due to the brane tensions. It is convenient to separate these distributional components from the continuous parts. It is further convenient to use separate bulk coordinates $\tensor*{x}{_n^\Gamma}$ in each bulk region \rn, rather than using a single global coordinate system. We will therefore have a bulk metric in each region \rn, rather than one global metric. We note that the $n^{\mathrm{th}}$ brane will then have two embedding functions: $\tensor*[^n]{x}{_n^\Gamma} (\tensor*{w}{_n^a})$ in the coordinates $\tensor*{x}{_n^\Gamma}$ of \rn, and $\tensor*[^n]{x}{_{n+1}^\Gamma}(\tensor*{w}{_n^a})$ in the coordinates $\tensor*{x}{_{n+1}^\Gamma}$ of ${\cal{R}}_{n+1}$.

Combining these modifications, we can write Eq. \eqref{generalaction} as
\begin{align}
S \left[ \tensor{g}{_{\Gamma \Sigma}}, \tensor[^n]{x}{^\Gamma}, \n\phi \right] ={}& \sum_{n = 0}^N \int_{\rn} d^5 \lrn{x} \sqrt{-\n g}
\left(\frac{\n[^{(5)}]R}{2 \kapfs} - \lrn \Lambda \right)+ \sum_{n = 0}^{N-1} \frac{1}{\kapfs} \int_{\bn} d^4 \lrn{w} \sqrt{-\n h} \left(\n[^+]K + \n[^-]K \right) \nonumber
\\
& {}- \sum_{n = 0}^{N-1} \lrn{\sigma}\int_{\bn} d^4 \lrn{w} \sqrt{-\n h} + \sum_{n = 0}^{N-1} \n[_m]S[\n[_{ab}]{h}, \n\phi] \label{coordinateAction}
\end{align}
where $\n[^+]{K}$ is the trace of the extrinsic curvature tensor of the $n^{\mathrm{th}}$ brane in the bulk region ${\cal{R}}_{n+1}$, and $\n[^-]{K}$ is the trace of the extrinsic curvature tensor of the $n^{\mathrm{th}}$ brane in the bulk region \rn, where the normals are always defined to be pointing away from the bulk region and towards the brane [see Eqs. \eqref{normal1} and \eqref{normal2} below]. These terms are just the usual Gibbons-Hawking terms \cite{Gibbons1977}.

A note on our conventions. Many functions, coordinates and parameters will be indexed by some index $n$ in this paper. For coordinates and parameters, the index will always be in the lower right, e.g, \lrn{x}. For functions, the index will be in the upper left, e.g, \n[_{\alpha \beta}]{g}. We use capital Greek letters ($\Gamma, \Sigma, \Theta$) to index five-dimensional tensors in arbitrary coordinate systems. When we specialize our coordinate system, we will use lowercase Greek letters ($\alpha, \beta, \gamma$) to index five-dimensional tensors. We use Roman letters $(a, b, c)$ for four-dimensional tensors. The metric $g$ refers to a five-dimensional metric, while the metric $h$ refers to a four-dimensional metric. Finally, we use the $(-++++)$ metric signature.

\subsection{Specializing the Coordinate System}
We begin by specializing the coordinate systems in each bulk region. Denote the coordinates by $\tensor*{x}{_n^\Gamma} = (\tensor*{x}{_n^a}, \lrn{y})$, where $a$ indicates one temporal and three spatial dimensions. Without loss of generality, we can choose the coordinates such that the branes bounding the region are located at fixed \lrn{y}. Next, choose the \lrn{y} coordinates such that the branes are located at $\lrn{y} = n-1$ and $\lrn{y} = n$. In other words, in the brane embedding functions $\tensor*[^n]{x}{^\Gamma_n}(\tensor*{w}{_n^a})$,
\begin{align}
\tensor[^{n-1}]{y}{_n}(\tensor*{w}{_{n-1}^a}) ={}& n-1,
\\
\tensor[^{n}]{y}{_n}(\tensor*{w}{_{n}^a}) ={}& n.
\end{align}
In this way, the first brane will be located at $\tensor{y}{_0}  = \tensor{y}{_1} = 0$, and the last brane located at $\tensor{y}{_{N-1}} = \tensor{y}{_N} = N-1$. The $n^{\mathrm{th}}$ bulk region \rn\ then extends from $\tensor{y}{_n} = n-1$ to $\tensor{y}{_n} = n$, with the exceptions of the first and last bulk regions, which extend away from the branes to $\mp \infty$ respectively.

Next, we use some of the available gauge freedom to remove off-diagonal elements of the metrics. Carena et al. \cite{Carena2005} have shown that it is always possible to find a coordinate transformation in \rn\ of the form $\tensor*{x}{_n^a} \rightarrow \tensor*{b}{_n^a} (\tensor*{x}{_n^c}, \lrn{y})$ to make $\n[_{ya}]{g} = 0$ while simultaneously maintaining that the branes be located at $\tensor{y}{_n} = n-1$ and $\tensor{y}{_n} = n$. After such a transformation, the metric in \rn\ can be written as
\begin{align}
\n[^2]{ds} ={}& \n[_{ab}]{\gamma}(\tensor*{x}{_n^c}, \lrn{y}) \tensor*{dx}{_n^a} \tensor*{dx}{_n^b} + \n[^2]{\Phi}(\tensor*{x}{_n^c}, \lrn{y}) \tensor*{dy}{_n^2} \label{firstmetric}
\end{align}
where the sign of $\n[_{yy}]{g}$ is known from the signature of the metric. We choose the sign of $\n\Phi$ to be positive.

The brane positions are now hyperplanes located at $y_n = \mathrm{integer}$. It is obvious that only coordinate transformations for which $y \rightarrow g(y)$ (with no $x^a$ dependence) can preserve this form for the hyperplanes. With this condition, only coordinate transformations for which $x^a \rightarrow f^a(x^b)$ will preserve the form of the metric. Thus, the remaining gauge freedom lies in coordinate transformations of the form $x^a \rightarrow f^a(x^b)$ and $y \rightarrow g(y)$ such that the positions of the branes are preserved.

For later simplicity, we choose the following parameterization of the four-dimensional metric $\n[_{ab}]\gamma$. In each bulk region, let
\begin{equation}
\n[_{a b}]\gamma (\tensor*{x}{^c_n}, y_n) = e^{\n\chi (\tensor*{x}{^c_n},
y_n)} \, \n[_{a b}]{\hat{\gamma}} (\tensor*{x}{^c_n}, y_n)
\end{equation}
such that the determinant of $\n[_{a b}]{\hat{\gamma}}$ is constrained to be $-1$. The function $\exp(\n\chi)$ is sometimes called the warp factor. The metric in \rn\ is then
\begin{equation}
\n[^2]{ds} = e^{\n\chi (\tensor*{x}{^c_n}, y_n)} \,\n[_{a b}]{\hat{\gamma}} (\tensor*{x}{^c_n}, y_n) dx_n^a dx_n^b + \n[^2]{\Phi} (\tensor*{x}{^c_n}, y_n) \tensor*{dy}{_n^2}. \label{metric}
\end{equation}

\subsection{Embedding Functions, Coordinate Systems on the Branes, and Induced Metrics}
We now specialize the coordinate system $\tensor*{w}{_n^a}$ on the $n^{\mathrm{th}}$ brane \bn. We choose the coordinate system on ${\cal{B}}_0$ to coincide with the first four coordinates of the bulk coordinate system of ${\cal{R}}_0$, evaluated on the brane. Thus,
\begin{subequations}
\begin{align}
\tensor*[^0]{x}{_0^\Gamma}(\tensor*{w}{_0^a}) ={}& (\tensor*[^0]{x}{_0^a}(\tensor*{w}{_0^a}), \tensor*[^0]{y}{_0}(\tensor*{w}{_0^a}))
\\
={}& (\tensor*{w}{_0^a}, 0).
\end{align}
Now, transform the coordinates in the second bulk region by transforming $\tensor*{x}{_1^a}$ such that
\begin{equation}
\tensor*[^0]{x}{_1^\Gamma}(\tensor*{w}{_0^a}) = (\tensor*{w}{^a_0}, 0).
\end{equation}
Such a transformation only requires a mapping of the form $\tensor*{x}{_1^a} \rightarrow f^a(\tensor*{x}{_1^b})$, and so the locations of the branes are preserved. Next, choose a coordinate system $\tensor*{w}{_1^a}$ on ${\cal{B}}_1$ such that
\begin{equation}
\tensor*[^1]{x}{_1^\Gamma}(\tensor*{w}{_1^a}) = (\tensor*{w}{^a_1}, 1)
\end{equation}
and continue this process until all branes and bulk regions have related coordinate systems. The coordinate systems we acquire have the property that for a point ${\cal P}$ on \bn, we have
\begin{equation}
\tensor*[^n]{x}{_{n}^\Gamma} ({\cal P}) = \tensor*[^n]{x}{_{n+1}^\Gamma} ({\cal P}). \label{CoordCondition}
\end{equation}
\end{subequations}
Note that while the condition \eqref{CoordCondition} implies that the coordinate patches can be joined continuously from one region to another in a straightforward manner, they need not form a global coordinate system as they may not join smoothly across the branes.

From the embedding functions in these coordinate systems we can calculate the induced metric on the branes, using Eq. \eqref{Embedding}. As each brane is adjacent to two bulk regions, there will be two induced metrics, one from each bulk region. For \bn, the induced metric from \rn\ is
\begin{align}
\n[_{ab}^-]{h}(\tensor*{w}{_n^c}) = e^{\n\chi (\tensor*{w}{^c_n}, n)} \,\n[_{a b}]{\hat{\gamma}} (\tensor*{w}{^c_n}, n)
\end{align}
while the induced metric from ${\cal{R}}_{n+1}$ is
\begin{align}
\n[_{ab}^+]{h}(\tensor*{w}{_n^c}) = e^{\np\chi (\tensor*{w}{^c_{n}}, n)} \,\np[_{a b}]{\hat{\gamma}} (\tensor*{w}{^c_{n}}, n)
\end{align}
We will restrict attention to configurations where the two induced metrics coincide (as would be enforced by the first Israel junction condition \cite{Israel1966}). We then have
\begin{align}
\n[_{ab}]{h}(\tensor*{w}{_n^c}) ={}& \n[_{ab}^-]{h}(\tensor*{w}{_n^c}) = \n[_{ab}^+]{h}(\tensor*{w}{_n^c})
\\
\n[_{ab}]{h}(\tensor*{w}{_n^c}) ={}& e^{\n\chi (\tensor*{w}{^c_n}, n)} \,\n[_{a b}]{\hat{\gamma}} (\tensor*{w}{^c_n}, n) = e^{\np\chi (\tensor*{w}{^c_n}, n)} \,\np[_{a b}]{\hat{\gamma}} (\tensor*{w}{^c_n}, n). \label{firstIsrael}
\end{align}
Taking the determinant of this expression and using the fact that the determinants of $\hat{\gamma}_{a b}$ are constrained to be $-1$, we find that
\begin{equation}
\n\chi (\tensor*{w}{^c_n}, n)= \np\chi (\tensor*{w}{^c_n}, n). \label{metricjunction} 
\end{equation}
Then by equation \eqref{firstIsrael}, it follows that
\begin{equation}
\n[_{a b}]{\hat{\gamma}} (\tensor*{w}{^c_n}, n) = \np[_{a b}]{\hat{\gamma}} (\tensor*{w}{^c_n}, n).
\label{firstIsraelgamma}
\end{equation}

\subsection{The Action}
Now that we have specialized the coordinate systems for every region and brane in our model, we can rewrite our action \eqref{coordinateAction} in terms of these coordinates.

We can evaluate the extrinsic curvature tensor terms as follows. Each brane has two normal vectors, one each from the two adjacent bulk regions. We define the normal vectors $\n[^\pm]{\vec{n}}$ at \bn\ to be the inward pointing normals from ${\cal{R}}_{n+1}$ and \rn. Since the branes are at fixed values of the coordinates \lrn{y}, this gives
\begin{equation}
\n[^-]{\vec{n}}(\tensor*{w}{_n^a}) = \frac{1}{\n\Phi (\tensor*{w}{_n^a}, n)} \partial_{\lrn y} \label{normal1}
\end{equation}
as the normal vector from \rn\, and
\begin{equation}
\n[^+]{\vec{n}}(\tensor*{w}{_n^a}) = -\frac{1}{\np\Phi (\tensor*{w}{_n^a}, n)} \partial_{\tensor*{y}{_{n+1}}} \label{normal2}
\end{equation}
as the normal vector from ${\cal{R}}_{n+1}$. The vector $\n[^-]{\vec{n}}$ points to the right of bulk region $n$ towards brane $n$, while $\n[^+]{\vec{n}}$ points to the left of region $n+1$ towards brane $n$.

For the extrinsic curvature tensors, we have by definition
\begin{align}
\n[_{ab}^-]K(\tensor*{w}{_n^c}) ={}& \left.
\frac{\partial (\tensor[^n]{x}{^\alpha})}{\partial \tensor*{w}{_n^a}}
\frac{\partial (\tensor[^n]{x}{^\beta})}{\partial \tensor*{w}{_n^b}}
\tensor{\nabla}{_\beta} \n[_\alpha^{-}]n \right|_{\tensor*{x}{^c_n} = \tensor*{w}{_n^c}, \lrn{y} = n} ,
\\
\n[_{ab}^+]K(\tensor*{w}{_n^c}) ={}& \left.
\frac{\partial (\tensor[^{n+1}]{x}{^\alpha})}{\partial \tensor*{w}{_n^a}}
\frac{\partial (\tensor[^{n+1}]{x}{^\beta})}{\partial \tensor*{w}{_n^b}}
\tensor{\nabla}{_\beta} \n[_\alpha^{+}]n \right|_{\tensor*{x}{^c_{n+1}} = \tensor*{w}{_n^c}, \tensor{y}{_{n+1}} = n}.
\end{align}
Evaluating these using the explicit form of the normals, we have
\begin{align}
\n[_{ab}^-]K(\tensor*{w}{_n^c}) ={}& \frac{1}{2} \frac{1}{\n \Phi} \left( \n[_{,y}]{\chi} e^{\n\chi} \,\n[_{ab}]{\hat{\gamma}} + e^{\n\chi} \,\n[_{ab,y}]{\hat{\gamma}} \right) (\tensor*{w}{_n^c}, n) ,
\\
\n[_{ab}^+]K(\tensor*{w}{_n^c}) ={}& -\frac{1}{2} \frac{1}{\np \Phi} \left( \np[_{,y}]{\chi} e^{\np\chi} \,\np[_{ab}]{\hat{\gamma}} + e^{\np\chi} \,\np[_{ab,y}]{\hat{\gamma}} \right) (\tensor*{w}{_n^c}, n).
\end{align}
To take the trace of the extrinsic curvature tensor, we contract with the inverse induced metric
\begin{equation}
\n[^{ab}]h = e^{-\n \chi} \,\n[^{ab}]{\hat{\gamma}} = e^{-\np \chi} \,\np[^{ab}]{\hat{\gamma}}.
\end{equation}
We find
\begin{align}
\n[^+]K (\tensor*{w}{_n^c}) 
={}& \left.-\frac{2 \;\np[_{,y}]{\chi}}{\np \Phi}\right|_{\tensor*{w}{_n^c}, n}, \label{ExtrinsicP}
\\
\n[^-]K (\tensor*{w}{_n^c}) ={}& \left.\frac{2 \;\n[_{,y}]\chi}{\n\Phi}\right|_{\tensor*{w}{_n^c}, n}. \label{ExtrinsicM}
\end{align}
In deriving these equations, we used the fact that $\n[^{ab}]{\hat{\gamma}} \; \n[_{ab, y}]{\hat{\gamma}} = 0$, which follows from $\det (\n[_{ab}]{\hat{\gamma}}) = -1$.


From Eq. \eqref{metric}, the determinant of the five-dimensional metric can be written as
\begin{align}
\sqrt{-\n g} = \n \Phi e^{2 \, \n \chi} \sqrt{-\n{\hat{\gamma}}}. \label{metricexpansion}
\end{align}
We do not substitute $\sqrt{-\n{\hat{\gamma}}} = 1$ at this stage; instead we choose to enforce this at the level of the action by a Lagrange multiplier (see Appendix \ref{AppExactEOMS}). Using Eqs. \eqref{ExtrinsicP}, \eqref{ExtrinsicM}, and \eqref{metricexpansion}, the action \eqref{coordinateAction} can be written as
\begin{align}
S \left[ \tensor[^n]{\hat{\gamma}}{_{ab}}, \n\Phi, \n\chi, \n\phi \right] ={} & \sum_{n = 0}^N \int_{\rn} d^5 \lrn x \n \Phi e^{2 \, \n \chi} \sqrt{-\n{\hat{\gamma}}}
\left(\frac{\n[^{(5)}]R}{2 \kapfs} - \lrn \Lambda\right) \nonumber
\\
& {} + \sum_{n = 0}^{N-1} \frac{2}{\kapfs} \int_{\bn} d^4 \lrn w e^{2 \, \n \chi(n)} \sqrt{-\n{\hat{\gamma}}} \left(\left.\frac{\n[_{,y}]\chi}{\n\Phi}\right|_{\lrn y = n} - \left.\frac{\np[_{,y}]\chi}{\np\Phi}\right|_{\tensor*{y}{_{n+1}} = n} \right) \nonumber
\\
& - \sum_{n = 0}^{N-1} \lrn \sigma \int_{\bn} d^4 \lrn w e^{2 \, \n \chi(n)} \sqrt{-\n{\hat{\gamma}}} + \sum_{n = 0}^{N-1} \n[_m]S [\n[_{ab}]h, \n \phi]. \label{CompleteAction}
\end{align}

%% file: lengthscale.tex
\section{Separation of Lengthscales\label{SecLengthscale}}
We now describe the approximation method, based on a two-lengthscale expansion, which we use to obtain a four-dimensional description of the system. We begin by defining the appropriate lengthscales, and then detail how the theory simplifies in the regime where the ratio of lengthscales is small.

\subsection{Two Lengthscales}
There are three groups of parameters in our model: the five-dimensional gravitational scale $\kappa_5^{2}$, the brane tensions $\{\sigma_n\}$, and the bulk cosmological constants $\{\Lambda_n\}$. We assume that all parameters in a group are of the same order of magnitude, and so will just consider typical parameters $\sigma$ and $\Lambda$. Working with units in which $c=1$, the dimensionality of these parameters in terms of mass units $M$ and length units $L$ are $[\kapfs] = {L^2}/{M}$, $[\sigma] = {M}/{L^3}$, and $[\Lambda] = {M}/{L^4}$.

We assume that the dimensionless parameter $\sigma^2 \kapfs/\Lambda$ is roughly of order unity; this will be enforced by the brane tuning conditions we derive below [see Eq. \eqref{branetunings}]. Eliminating \kapfs, we can then define a lengthscale by
\begin{align}
{\cal L} = {\sigma}/{\Lambda} \label{defL}
\end{align}
and a mass scale by
\begin{align}
{\cal M} = {\sigma^4}/{\Lambda^3}. \label{defM}
\end{align}

For a given configuration, we also define a \textit{four-dimensional curvature lengthscale} ${\cal L}_c(y)$ on each slice of constant $y$, as follows. We take the minimum of the transverse lengthscale over which the induced metric varies, and the transverse lengthscale over which the metric coefficient $\Phi$ varies. In other words,
\begin{align}
{\cal L}_c (y) \sim \min \left\{ \left|\tensor*{R}{^{(4)}_{\hat{a} \hat{b} \hat{c} \hat{d}}}\right|^{-1/2}, \left|\tensor*{\nabla}{_{\hat{a}}} \tensor*{R}{^{(4)}_{\hat{b} \hat{c} \hat{d} \hat{e}}}\right|^{-1/3}, \ldots, \frac{\left|\Phi\right|}{\left|\tensor*{\nabla}{_{\hat{a}}} \Phi\right|}, \frac{\left|\Phi\right|^{1/2}}{\left|\tensor*{\nabla}{_{\hat{a}}} \tensor*{\nabla}{_{\hat{b}}} \Phi\right|^{1/2}}, \ldots \right\}
\end{align}
where $\hat{a}, \hat{b}, \ldots$ denotes an orthonormal basis of the induced metric, $R_{\hat{a} \hat{b} \hat{c} \hat{d}}$ is the Riemann tensor of the induced metric, and dots denote similar terms with more derivatives.

Thus, for a given configuration, we have two natural lengthscales: the microphysical lengthscale ${\cal L} = \sigma / \Lambda$ (the same for all configurations), and the macrophysical curvature lengthscale ${\cal L}_c$ (where the $c$ is intended to denote ``curvature'').

\subsection{Separating the Lengthscales}
We now evaluate the action \eqref{CompleteAction} in the low energy regime ${\cal L}_c \gg {\cal L}$, in which the theory admits a four-dimensional description. We will find that there is a leading order term of order $\sim {\cal ML}$, and a subleading term of order $\sim {\cal ML} ({\cal L}/{\cal L}_c)^2$. Our strategy will be to separate the contributions to the action at each order, minimize the leading order piece of the action, and then substitute the general solutions obtained from that minimization into the subleading piece of the action. The result will be a four-dimensional action which gives the effective description of the system in the low energy regime.

We write the action \eqref{CompleteAction} as a sum $S = S_g + S_m$ of a gravitational part $S_g$ and a matter part $S_m$, where the matter part is the last term in Eq. \eqref{CompleteAction} and the gravitational part comprises the remaining terms.

We first discuss the expansion of the gravitational action $S_g$, which is a functional of a bulk metric $\tensor{g}{_{\alpha \beta}}$ and brane embedding functions $\tensor[^n]{x}{^\Gamma}$. We define a mapping $T_\epsilon$ which acts on these variables
\begin{align}
T_\epsilon : (\tensor{g}{_{\alpha \beta}}, \tensor[^n]{x}{^\Gamma}) \rightarrow (\tensor*{g}{_{\alpha \beta}^\epsilon}, \tensor*[^n]{x}{^\Gamma_\epsilon}),
\end{align}
where $\epsilon > 0$ is a dimensionless parameter, as follows: (i) We specialize to our chosen gauge, (ii) replace the metric \eqref{metric} with the rescaled version
\begin{equation}
\tensor*{ds}{^2_\epsilon} = \frac{1}{\epsilon^2} e^{\chi(x^c, y)} \tensor{\hat{\gamma}}{_{a b}} (x^c, y) dx^{a} dx^{b} + \Phi^{2} (x^c, y) dy^2 \label{scaledmetric},
\end{equation}
where indices indicating regions have been suppressed, and (iii) leave the embedding functions in our chosen gauge unaltered. We may think of $\epsilon$ as a parameter which tunes the ratio of the microphysical lengthscale to the macrophysical lengthscale. As $\epsilon$ is decreased, lengthscales on the brane are inflated, and so ${\cal L}_c$ increases. Thus, as $\epsilon$ decreases, so does the ratio ${\cal L} / {\cal L}_c$. In particular, we have
\begin{align}
\left(\frac{{\cal L}}{{\cal L}_c}\right)_\epsilon = \epsilon \frac{{\cal L}}{{\cal L}_c}.
\end{align}
It is important to note that this $\epsilon$ scaling does not map solutions to solutions, but just provides a means of keeping track of the dependence on the various lengthscales.

We can construct a one-parameter family of action functionals by using these rescaled metrics in our original action \eqref{CompleteAction}\footnote{The factor of $\epsilon^4$ in Eq. \eqref{epsilonaction} is inserted for convenience, so that Eq. \eqref{scaledaction} contains terms of $O(1)$ and $O(\epsilon^2)$. This is explicitly shown in Section \ref{SecLowest}.}:
\begin{equation}
\tensor{S}{_{g, \epsilon}} \left[\tensor*{g}{_{\alpha \beta}}, \n[^\Gamma]x \right] \equiv \epsilon^4 S_{g} \left[\tensor*{g}{_{\alpha \beta}^\epsilon}, \n[^\Gamma_\epsilon]x \right]. \label{epsilonaction}
\end{equation}
We can expand this action in powers of $\epsilon$ by
\begin{equation}
\tensor{S}{_{g, \epsilon}} \left[\tensor*{g}{_{\alpha \beta}}, \n[^\Gamma]x \right] = \tensor{S}{_{g, 0}}\left[\tensor*{g}{_{\alpha \beta}}\right] + \epsilon^2 \tensor{S}{_{g, 2}}\left[\tensor*{g}{_{\alpha \beta}}\right], \label{scaledaction}
\end{equation}
where on the right hand side we omit the dependence on the embedding functions since we have used the gauge freedom to fix those. The expansion \eqref{scaledaction} truncates after 2 terms; there are no higher order terms in $\epsilon$. Note that there is no $O(\epsilon)$ term, as when the action \eqref{epsilonaction} is evaluated, terms of $O(\epsilon^2)$ arise from contractions in the Ricci scalar using $g^{ab}$ ($O(1)$ terms arise from $g^{yy}$ contractions). Terms of order $O(\epsilon)$ would arise from contractions using $g^{ay}$, but as these components of the metric have been gauge-fixed to zero, they are not present. This can be seen explicitly in the expansion of the Ricci scalar \eqref{Ricci2expansion}. As we tune $\epsilon \rightarrow 0$, we move further into the low energy regime, and so we identify the zeroth order term as the dominant contribution to the action, and the second order term as the subleading term. This provides the separation of lengthscales we desire.


Let us now turn to the matter contribution to the action, $S_m$. We expect the matter action to contribute at $O(\epsilon^2)$, the same order as the subleading gravitational term. To see this, note that the brane tensions scales as $\sigma \sim {\cal M}/{\cal L}^3$, where the scales ${\cal M}$ and ${\cal L}$ were defined in Eqs. \eqref{defL} and \eqref{defM}. The matter action will be roughly $S_m \sim \int \rho \; d^4 x$, where $\rho$ is a 4-dimensional energy density. The four-dimensional Newton constant $\kappa_4^2 = 8 \pi G$ is of order $\kappa_4^2 \sim {\cal L} / {\cal M}$ by dimensional analysis [c.f. Eq. \eqref{4dNewtonConst} below], and so $\rho$ will be of order
\begin{align}
\rho \sim \frac{1}{\kappa_4^2 {\cal L}_c^2} \sim \frac{{\cal M}}{{\cal L} {\cal L}_c^2}.
\end{align}
Taking the ratio $\rho / \sigma$ now gives
\begin{align}
\frac{\rho}{\sigma} \sim \frac{{\cal M} / {\cal L} {\cal L}_c^2}{{\cal M} / {\cal L}^3} \sim \frac{{\cal L}^2}{{\cal L}_c^2} \propto \epsilon^2. \label{rhoscaling}
\end{align}
Formally, the scaling \eqref{rhoscaling} can be achieved by replacing the matter action $S_m$ with a rescaled action $S_{m, \epsilon}$ given by (i) multiplying by $\epsilon^4$ as in Eq. \eqref{epsilonaction}, (ii) rescaling all fields and dimensional constants with dimensions $(\mathrm{mass})^r (\mathrm{length})^s$ by factors of $\epsilon^{-(r+s)}$. The expansion of the full action is then
\begin{align}
S_\epsilon = S_{g, \epsilon} + S_{m, \epsilon} = {}& S_{g, 0} + \epsilon^2 \left[ S_{g, 2} + S_m \right] \nonumber
\\
= {}& S_0 + \epsilon^2 S_2. \label{completeepsilonaction}
\end{align}
It can be seen that given brane tensions tuned to the bulk cosmological constants, $\sigma^2 \sim \Lambda/\kapfs$, we require that the matter density on a brane should be small, so as not to spoil the tuning. This also yields $\rho \ll \sigma$, which roughly corresponds to the separation of lengthscales condition ${\cal L} \ll {\cal L}_c$.

We perform this $\epsilon$ scaling separately in each bulk region of the model. The contribution to the action from each region will separate into zeroth and second order terms.

\subsection{The Low Energy Regime\label{secdiscuss}}
Now that the contributions to each order have been identified, we can minimize the leading order term in the action, $\tensor{S}{_0}$. Once general solutions to the equations of motion have been found, we can use these solutions in the second order term in the action. Thus, we solve for the high energy (short lengthscale) dynamics first, and use the solution to this as a background solution for the low energy (long lengthscale) dynamics. At this point, we may let $\epsilon \rightarrow 1$, and rely on the ratio $({\cal L}/{\cal L}_c)^2$ being sufficiently small to provide the separation of lengthscales.

The effect of this separation of lengthscales is to enforce a decoupling of the high energy dynamics from the low energy dynamics. We will see below that the equation of motion for the high energy dynamics contains $y$ derivatives, but no $x^a$ derivatives. The theory at this order thus reduces to a set of uncoupled theories, one along each fiber $x^a = \mathrm{const}$ in the bulk. These theories are coupled together at $O(\epsilon^2)$, and thus in the regime of interest, the coupling is minimal. After solving the high energy dynamics along these fibres, a four-dimensional effective description of the system remains.

The low energy regime, in which the theory admits a four-dimensional description, is the regime
\begin{align}
{\cal L}_c \gg {\cal L}. \label{comparisonofls}
\end{align}
This regime is also frequently characterized in the literature by the condition
\begin{align}
\rho \ll \sigma, \label{sigmarho}
\end{align}
where $\rho$ is the mass density on a brane and $\sigma$ is a brane tension [c.f. Eq. \eqref{rhoscaling} above]. One can interpret the condition \eqref{sigmarho} as saying that the mass density on the brane must be sufficiently small that the brane-tuning conditions [Eq. \eqref{branetunings} below which enforces $\sigma^2 \sim \Lambda / \kapfs$] are not appreciably modified. However, the condition \eqref{sigmarho} is less general than the condition \eqref{comparisonofls}, and although necessary, is actually insufficient. First, as discussed above, \eqref{sigmarho} only applies to branes, whereas \eqref{comparisonofls} applies at each value of $y$, including away from the branes. Second, even when the density on a given brane vanishes, four-dimensional gravitational waves on that brane can give rise to radii of curvature ${\cal L}_c$ that are comparable to ${\cal L}$. In this case, the separation of lengthscales will not apply and the four-dimensional effective theory will not be valid, despite the fact that the condition \eqref{sigmarho} is satisfied. Curvature associated with the metric coefficient $\Phi$ can also yield similar results.

Finally, we discuss a subtlety in our definition of the ``low energy regime''. As noted in the previous paragraph, ${\cal L}_c$ varies with position in the five-dimensional universe. Our separation of lengthscales will break down when the induced metric on any slice of constant $y$ has a radius of curvature ${\cal L}_c$ comparable to that of the microphysical lengthscale ${\cal L}$; it is insufficient to require that ${\cal L} \gg {\cal L}$ on each brane. When this happens, the terms of order $\epsilon^2$ will couple strongly to the $O(1)$ terms, and our approximate solutions for the five-dimensional metric will no longer be valid. This will generically occur at sufficiently large distances from the branes, as $\exp(\n\chi)$ typically grows exponentially small away from the branes, and ${\cal L}_c^{-2} \propto \exp(-\n\chi) R^{(4)}$. Despite this breakdown, the contribution to the action from these regimes is exponentially suppressed by the warp factor, and thus provides only a small deviation from the effective theory. It is unlikely that the warp factor can grow without bound after encountering this regime while maintaining a globally hyperbolic spacetime.

%% file: 0thorder.tex
\section{The Action to Lowest Order\label{SecLowest}}
In this section, we calculate and explicitly solve the equations of motion to lowest order in the two lengthscale expansion. First, however, we write out the complete, rescaled action showing explicitly the dependence on $\epsilon$.  Inserting the decomposition \eqref{Ricci2expansion} of the Ricci scalar and the rescaled metric \eqref{scaledmetric} into the action \eqref{CompleteAction} [following the prescription of Eq. \eqref{epsilonaction}], we obtain
\begin{align}
S_\epsilon \left[ \tensor[^n]{\hat{\gamma}}{_{ab}}, \n\Phi, \n\chi, \n\phi \right] ={} &
\sum_{n = 0}^N \int_{\rn} d^5 \lrn x \Bigg[ \sqrt{-\n{\hat{\gamma}}} \frac{e^{2 \, \n\chi }}{2 \kapfs \, \n\Phi } \bigg( - \frac{1}{4}\n[^{ab}]{\hat{\gamma}} \n[_{bc,y}]{\hat{\gamma}} \n[^{cd}]{\hat{\gamma}} \n[_{da,y}]{\hat{\gamma}} - 5 (\n[_{,y}]\chi)^2 - 4 \n[_{,yy}]\chi \nonumber
\\
{}& + 4 \frac{\n[_{,y}]\Phi}{\n\Phi} \n[_{,y}]\chi - 2 \kapfs \, \n[^2]\Phi \lrn \Lambda \bigg) + \n\lambda (x^a, y) \left(\sqrt{-\n{\hat{\gamma}}} - 1\right) \Bigg] \nonumber
\\
{}& {} + \sum_{n = 0}^{N-1} \int_{\bn} d^4 \lrn w e^{2 \, \n\chi(n)} \sqrt{-\n{\hat{\gamma}}} \left[ \frac{2}{\kapfs} \left(\left.\frac{\n[_{,y}]\chi}{\n\Phi}\right|_{\lrn y = n} - \left.\frac{\np[_{,y}]\chi}{\np\Phi}\right|_{\tensor*{y}{_{n+1}} = n} \right) - \lrn \sigma \right] \nonumber
\\
{}& + \sum_{n = 0}^N \epsilon^2 \int_{\rn} d^5 \lrn x \sqrt{-\n{\hat{\gamma}}} \frac{e^{\n \chi}}{2 \kapfs} \bigg(\n \Phi \n[^{(4)}]R - 3 \n \Phi \nabla^a \nabla_a \n\chi  - \frac{3}{2} \n \Phi (\nabla^a \n\chi) (\nabla_a \n\chi) \nonumber
\\
{}& - 2 \nabla^a \nabla_a \n\Phi - 2 (\nabla^a \n\chi) (\nabla_a \n\Phi) \bigg) + \sum_{n = 0}^{N-1} \epsilon^2 \, \n[_m]S [\n[_{ab}]h, \n \phi] \label{scaledepsilonaction}
\end{align}
where we include the Lagrange multiplier terms \eqref{Lagrangemults} discussed in Appendix \ref{AppExactEOMS}, and where the factor of $\epsilon^2$ in front of the matter action comes from the process described in the previous section. This form explains the choice of the $\epsilon^4$ factor in Eq. \eqref{epsilonaction}, and shows the decomposition into $O(1)$ and $O(\epsilon^2)$ terms, as claimed in Eq. \eqref{completeepsilonaction}.

From the form of Eq. \eqref{scaledepsilonaction}, we see that we can neglect the last two lines in the limit $\epsilon \rightarrow 0$. We can obtain a more precise characterization of the domain of validity of this low energy approximation by estimating the ratio between the terms dropped and the terms retained. As an example, consider the first term on the $4^{\mathrm{th}}$ line and the first term on the first line. Their ratio is (dropping the `$n$' labels)
\begin{align}
\left[e^\chi \Phi R^{(4)}\right] \left[ \frac{e^{2\chi}}{\Phi} {\hat{\gamma}}^{ab} {\hat{\gamma}}_{bc,y} {\hat{\gamma}}^{cd} {\hat{\gamma}}_{da,y} \right]^{-1} \sim {}& \left[ e^{\chi} \Phi R^{(4)} \right] \left[ \frac{e^{2 \chi}}{\Phi \tilde{y}^2} \right]^{-1} \nonumber
\\
\sim {}& \left[ e^{-\chi} R^{(4)} \right] \left[ \Phi^2 \tilde{y}^2 \right]
\end{align}
where $\tilde{y}$ is the coordinate lengthscale over which $\hat{\gamma}_{ab}$ varies. We recognize the first factor as essentially the Ricci scalar of the induced metric $e^\chi \hat{\gamma}_{ab}$, which is of order ${\cal L}_{c}^{-2}$. We recognize the second factor as the square of the physical lengthscale in the $y$ direction over which $\hat{\gamma}$ varies, which is always $\sim {\cal L}^2$ (see the explicit solution \eqref{metricansatz} below). Thus, the ratio is $({\cal L}/{\cal L}_c)^2$, confirming the identification of the low energy regime as ${\cal L} \ll {\cal L}_c$.

\subsection{Varying the Action}
In the action \eqref{scaledepsilonaction} at zeroth order in $\epsilon$, we have three fields to vary (in $N$ regions): $\n\Phi (x^c, y)$, $\n\chi (x^c, y)$, and $\n[_{ab}]{\hat{\gamma}} (x^c, y)$. There is a subtlety in the variation however. The constraint that $\det \left(\n[_{ab}]{\hat{\gamma}}\right) = -1$ must be imposed either at the level of the equations of motion, or by a Lagrange multiplier. The Lagrange multiplier is explicitly included in Eq. \eqref{scaledepsilonaction}. Further details are provided in Appendix \ref{AppExactEOMS}.

We begin by varying with respect to $\n\Phi$. From this variation, we find a single equation of motion in each region,
\begin{align}
\frac{1}{4} \n[^{ab}]{\hat{\gamma}} \; \n[_{bc, y}]{\hat{\gamma}} \n[^{cd}]{\hat{\gamma}} \; \n[_{da, y}]{\hat{\gamma}} - 3 \; \n[_{, y}^2]\chi - 2 \kapfs \n[^2]\Phi \lrn{\Lambda} = 0 . \label{eqmotion2}
\end{align}

Next, we vary with respect to $\n[_{ab}]{\hat{\gamma}}$. Note that in varying the action, we obtain boundary terms from neighboring regions from the relationship \eqref{firstIsraelgamma}. The variation produces an equation of motion in each bulk region,
\begin{align}
\n[_{ad, yy}]{\hat{\gamma}} = \n[_{ab,y}]{\hat{\gamma}} \; \n[^{bc}]{\hat{\gamma}} \; \n[_{cd, y}]{\hat{\gamma}} - \n[_{ad, y}]{\hat{\gamma}} \left( 2 \; \n[_{,y}]{\chi} - \frac{\n[_{,y}]\Phi}{\n\Phi}\right). \label{eqYuck}
\end{align}
(If using Lagrange multipliers, this equation results after the Lagrange multiplier is eliminated by tracing the equation using \n[^{ab}]{\hat{\gamma}} and back substituting). Note that tracing over the indices and using Eq. \eqref{gammamotion} leads to Eq. \eqref{gammaswaps} as expected. We also find a boundary condition to be satisfied at each brane,
\begin{align}
\frac{1}{\n\Phi} \n[_{ab, y}]{\hat{\gamma}} (y_{n} = n) ={}& \frac{1}{\np\Phi} \np[_{ab, y}]{\hat{\gamma}} (y_{n+1} = n). \label{boundarygamma}
\end{align}

Finally, we vary with respect to $\n \chi$. Here, we once again have boundary terms arising from integrating bulk terms by parts in neighboring regions.
There is an equation of motion in each bulk region,
\begin{equation}
\frac{1}{12} \n[^{ab}]{\hat{\gamma}} \; \n[_{bc, y}]{\hat{\gamma}} \n[^{cd}]{\hat{\gamma}} \; \n[_{da, y}]{\hat{\gamma}} + \n[_{, y}^2]\chi + \n[_{, yy}]\chi - \frac{\n[_{,y}]\Phi}{\n\Phi} \n[_{,y}]\chi + \frac{2}{3} \kapfs \n[^2]\Phi \lrn{\Lambda} = 0. \label{eqmotion1}
\end{equation}
We also find a boundary condition at each brane,
\begin{equation}
\left.\frac{\n[_{,y}]\chi}{\n\Phi}\right|_{\lrn y = n} - \left.\frac{\np[_{,y}]\chi}{\np\Phi}\right|_{\lrnp y = n} = \frac{2}{3} \kapfs \lrn{\sigma}. \label{jumpconditions}
\end{equation}

\subsection{Solving the Equations of Motion\label{eqmotions}}
We have three equations of motion for each bulk region, as well as numerous boundary conditions for the fields at the branes [Eqs. \eqref{metricjunction}, \eqref{firstIsraelgamma}, \eqref{eqmotion2}, \eqref{eqYuck}, \eqref{boundarygamma}, \eqref{eqmotion1}, and \eqref{jumpconditions}]. Note that these equations all describe the dynamics along a fibre of constant $x^a$ which doesn't couple to any other fibres, and so solving these equations of motion consists of solving the dynamics of the extra dimension of the model.

We begin by solving Eq. \eqref{eqYuck}. It is convenient to solve this equation in matrix notation. Let
\begin{align}
[{\hat{\gamma}}_{ab}] = \boldsymbol{\hat{\gamma}}
\end{align}
where we suppress indices $n$. Then in matrix notation, Eq. \eqref{eqYuck} is
\begin{align}
\boldsymbol{\ddot{\hat{\gamma}}} = \boldsymbol{\dot{\hat{\gamma}}} \ \boldsymbol{\hat{\gamma}}^{-1} \boldsymbol{\dot{\hat{\gamma}}} - \boldsymbol{\dot{\hat{\gamma}}} \left( 2 \chi_{,y} - \frac{\Phi_{,y}}{\Phi}\right),
\end{align}
where dots denote derivatives with respect to $y$. It is straightforward to check that a solution to this differential equation in region $n$ is
\begin{align}
\boldsymbol{\hat{\gamma}} (x^a, y) ={}& \mathbf{A}(x^a) \exp\left( \mathbf{B}(x^a) \int^y_{n-1} \Phi(x^a, y^\prime) e^{-2\chi(x^a, y^\prime)} dy^\prime \right). \label{gammasoln}
\end{align}
where $\mathbf{A}$ and $\mathbf{B}$ are arbitrary $4 \times 4$ real matrix functions of $x^a$. The lower limit on the integral is chosen so that the boundary conditions may be matched at the previous brane (obviously, care must be taken in the first region). The expression \eqref{gammasoln} has the correct number of integration constants to satisfy arbitrary boundary conditions. From our knowledge of $\hat{\gamma}_{ab}$, $\mathbf{A}$ must be a symmetric matrix with determinant $-1$. The exponential has unit determinant, and so $\mathbf{B}$ must be traceless. The condition that $\boldsymbol{\hat{\gamma}}$ is symmetric implies that $\mathbf{B}^T = \mathbf{A} \ \mathbf{B} \ \mathbf{A}^{-1}$. The quantity which appears in Eqs. \eqref{eqmotion2} and \eqref{eqmotion1} is
\begin{align}
\n[^{ab}]{\hat{\gamma}} \; \n[_{bc, y}]{\hat{\gamma}} \n[^{cd}]{\hat{\gamma}} \; \n[_{da, y}]{\hat{\gamma}} ={}& \n[^{ab}]{\gamma} \; \n[_{ab, yy}]{\gamma} \nonumber
\\
={}& \mathrm{Tr} \left( \mathbf{B}^2 (x^a) \right) \Phi^2 e^{-4 \chi}.
\end{align}
We define
\begin{align}
b(x^a) = \frac{1}{12} \mathrm{Tr} \left( \mathbf{B}^2 (x^a) \right)
\end{align}
where the factor of 12 has been chosen for later convenience. 
From combining Eq. \eqref{boundarygamma} with Eqs. \eqref{metricjunction} and \eqref{firstIsraelgamma}, we see that $\mathbf{B}$ (and thus $b(x^a)$) is independent of region, while $\mathbf{A}$ will change with each region according to Eq. \eqref{firstIsraelgamma}.

From Eq. \eqref{eqmotion2}, we find
\begin{align}
\n[_{,y}]\chi ={}& \pm \sqrt{b \n[^2]\Phi \exp(-4 \n\chi) - \frac{2}{3} \kapfs \n[^2]\Phi \lrn \Lambda} \nonumber
\\
={}& P_n \n\Phi \sqrt{b \exp(-4 \n\chi)- \frac{2}{3} \kapfs \lrn \Lambda} \label{eqchi}
\end{align}
where $P_n$ is either $\pm 1$ and is constant in each bulk region. Differentiating Eq. \eqref{eqchi} gives
\begin{align}
\n[_{,yy}]\chi ={}& \frac{\n[_y]\Phi}{\Phi} \chi_{,y} - 2 b \, \n[^2]\Phi e^{-4 \n\chi}.
\end{align}
The same result is obtained by substituting Eq. \eqref{eqmotion2} into Eq. \eqref{eqmotion1}, and so we see that these equations of motion are degenerate. This leaves only one equation of motion (Eq. \eqref{eqchi}) and one boundary condition (Eq. \eqref{jumpconditions}) to satisfy.

\subsection{Classes of Solutions\label{SecClasses}}
If $\mathbf{B}(x^a) \equiv \mathbf{0}$, then the induced metric on all the branes are related to one another by conformal transformations, and a four-dimensional effective action is easily calculated. On the other hand, when $\mathbf{B}(x^a) \neq \mathbf{0}$, the induced metrics on each brane are not simply related conformally, but through the equations \eqref{firstIsraelgamma} and \eqref{gammasoln}. If solutions with $\mathbf{B}(x^a) \neq \mathbf{0}$ were to exist, the four-dimensional effective theory would contain more than one massless tensor degree of freedom; i.e., it would constitute a multigravity theory (see Damour and Kogan \cite{Damour2002}). No such degrees of freedom have been seen in any linearized analyses\footnote{In addition, it can be shown that in orbifolded models, there are no solutions with $\mathbf{B}(x^a) \neq \mathbf{0}$; see Section \ref{secRSI}.}. It is important to note that this is not a Kaluza-Klein mode. We believe that solutions with $\mathbf{B}(x^a) \neq \mathbf{0}$ are ruled out due to divergences at $y \rightarrow \pm \infty$, leading to a lack of global hyperbolicity in the spacetime, although we have been unable to prove this rigorously. We will restrict attention to the case $\mathbf{B}(x^a) = \mathbf{0}$ for the remainder of this paper.

\subsection{General Solutions at Leading Order\label{BraneTensionDerivation}}
With $\mathbf{B}(x^a) \equiv \mathbf{0}$, the field $\n[_{ab}]{\hat{\gamma}}$ becomes independent of $y$ [see Eq. \eqref{gammasoln}], and also independent of $n$ by Eq. \eqref{firstIsraelgamma}. This means that we can drop the index $n$ from $x^a_n$, $w_n$, and $\n[_{ab}]{\hat{\gamma}}$ without causing confusion. With $b(x^a) = 0$, the remaining equation of motion and boundary condition simplify somewhat. Equation \eqref{eqchi} becomes
\begin{align}
\n[_{,y}]\chi ={}& P_n \n\Phi \sqrt{- \frac{2}{3} \kapfs \lrn \Lambda}, \label{neweqmotion}
\end{align}
which implies that $\lrn \Lambda < 0$, and so the bulk regions must be slices of anti de-Sitter space. Define
\begin{align}
k_n = \sqrt{\frac{-\kapfs \lrn \Lambda}{6}}. \label{kndef}
\end{align}
We can use Eq. \eqref{neweqmotion} for $\n\chi$ in Eq. \eqref{jumpconditions} to obtain
\begin{equation}
\lrn{k} \lrn{P} - \lrnp{k} \lrnp{P} = \frac{1}{3} \kapfs \lrn \sigma. \label{branetunings}
\end{equation}
These relations are the well-known ``brane tunings'', which determine the branes tensions required in order to avoid a cosmological constant on the branes \cite{Randall1999}.

We may integrate Eq. \eqref{neweqmotion} to find
\begin{align}
\n\chi(x^a, y) =
\begin{cases}
2 k_0 P_0 \int_0^y \n\Phi(x^a, y^\prime) dy^\prime + f(x^a) & n = 0,
\\
\nm\chi(x^a, n-1) + 2 \lrn{k} \lrn{P} \int_{n-1}^y \n\Phi(x^a, y^\prime) dy^\prime & n > 0
\end{cases} \label{chidef}
\end{align}
where $f(x^a)$ is an arbitrary function. Note that the field $\n\chi$ is related to the distance from the previous brane to $y$ along a geodesic normal to the branes, made dimensionless by the appropriate lengthscale in the bulk. In particular, $\chi$ describes the number of $e$-foldings the warp factor in the metric provides between two points in the five-dimensional spacetime. Assuming that $\Phi$ is not divergent, if $\exp(\n\chi(y))$ approaches zero or $\infty$ anywhere, it can only occur as $y \rightarrow \pm \infty$. We will restrict attention to the cases
\begin{align}
P_0 = +1, \ \ \mathrm{and} \ \ P_N = -1. \label{boundaryp}
\end{align}
When these signs fail to hold, then the warp factor increases monotonically as one goes to infinity, and it seems likely that the spacetime cannot be globally hyperbolic. We exclude cases where $\exp(\n\chi(y)) \rightarrow 0$ at finite $y$ by restricting ourselves to topologically connected spacetimes \cite{Flanagan2001a, Karch2001}.

%% file: wherenow.tex
\subsection{Summary}
We summarize our results so far. We have $N$ branes, each with a brane tension which has been carefully adjusted, according to Eq. \eqref{branetunings}. The branes divide our system into $N+1$ regions. Our coordinates are $x^\alpha$, describing four-dimensional space, and $y$, describing the extra dimension.

We expanded the action in terms of our $\epsilon$ scaling parameter to separate the high and low energy contributions. Specializing to a low energy regime, we solved for the high energy dynamics, arriving at the metric for each region of our system:
\begin{align}
\n[^2]{ds} = e^{\n \chi(x^c, y)} \hat{\gamma}_{ab} (x^c) dx^a dx^b + \frac{\n[^2_{,y}]\chi (x^c, y)}{4 k_n^2} dy^2, \label{metricansatz}
\end{align}
with $\n\chi$ given by Eq. \eqref{chidef}, where $\n\Phi(x^a, y)$ can be chosen freely. The parameters $k_n$ are determined by the bulk cosmological constants and the five-dimensional Newton's constant, by Eq. \eqref{kndef}. The derivative $\chi_{,y}$ has fixed sign $P_n = \pm 1$ in each region, although the derivative may approach zero as $y \rightarrow \pm \infty$.

As an aside, when the metric in each region is in the form \eqref{metricansatz}, the zeroth order action $S_0 [g_{ab}]$ [Eq. \eqref{scaledaction}] evaluates to exactly zero. This can be seen by substituting the metric \eqref{metricansatz} into the action and explicitly evaluating the integral over the $y$ dimension. All of the integrals become total derivatives whose boundary terms exactly cancel the boundary terms present in the action at this order.

The background metric ansatz \eqref{metricansatz} is essentially the same as the zeroth order metric calculated by Kanno and Soda \cite{Kanno2002}, taking $\Phi^2(x^a, y) = \exp(2 \phi (y, x))$ in their notation. However, from here, we proceed without their assumption that $\phi(y, x) = \phi(x)$. The ``na\"{\i}ve'' ansatz and the CGR ansatz of Chiba \cite{Chiba2000} are also in the form of our metric \eqref{metricansatz}.

%% file: 2ndorder.tex
\section{The Action to Second Order\label{SecSecond}}
In this section, we investigate the action to second order in $\epsilon$. By integrating out the previously determined high energy dynamics, we find the four-dimensional effective action.

\subsection{Acquiring the Four-Dimensional Effective Action\label{SecSecond1}}
Using the metric \eqref{metricansatz} in Eqs. \eqref{CompleteAction} and \eqref{scaledaction}, we can calculate the second order contribution to the action $S_2$. The result is
\begin{align}
 S_2 \left[ \tensor[^n]{\hat{\gamma}}{_{ab}}, \n\chi, \n\phi \right] = \sum_{n = 0}^N \int_{\rn} d^5 & \lrn x \sqrt{-\hat{\gamma}} \frac{e^{\n\chi}}{4 \kapfs \lrn k \lrn P} \Big[ \n[_{,y}]\chi R^{(4)} - 3 \; \n[_{,y}]\chi \nabla^2 \n\chi - \frac{3}{2} \n[_{,y}]\chi (\nabla^a \n\chi) (\nabla_a \n\chi) \nonumber
\\
 & \qquad {}- 2 \nabla^2 \n[_{,y}]\chi - 2 (\nabla^a \n\chi) (\nabla_a \n[_{,y}]\chi) \Big] + \sum_{n = 0}^{N-1} \n[_m]S \left[ e^{\n\chi (x^a, n)} \hat{\gamma}_{ab}, \n\phi \right]. \label{s2action}
\end{align}
Note that covariant derivatives as written here are associated with the metric $\hat{\gamma}_{ab}$, as is the four-dimensional Ricci scalar $R^{(4)}$.

To obtain the effective four-dimensional action, we integrate over $y$ in the five-dimensional action \eqref{s2action}, as the dynamics of this dimension have already been solved. The term involving the Ricci scalar can be integrated straightforwardly, as $R^{(4)}$ has no $y$ dependence, but the other terms require more manipulation. We can combine the last four terms in the five-dimensional integral in the following way:
\begin{align}
 -3 e^{\n\chi} \; \n[_{,y}]\chi \nabla^2 \n\chi & - \frac{3}{2} e^{\n\chi} \; \n[_{,y}]\chi (\nabla^a \n\chi) (\nabla_a \n\chi) - 2 e^{\n\chi} \nabla^2 \n[_{,y}]\chi - 2 e^{\n\chi} (\nabla^a \n\chi) (\nabla_a \n[_{,y}]\chi) \nonumber
\\
 ={}& \frac{3}{2} \frac{\partial}{\partial y} \left( e^{\n\chi} (\nabla^a \n\chi) (\nabla_a \n\chi) \right) - \nabla^a \left( 3 e^{\n\chi} \; \n[_{,y}]\chi \nabla_a \n\chi + 2 e^{\n\chi} \nabla_a \n[_{,y}]\chi \right)
\end{align}
The covariant derivative commutes with the integration over the fifth dimension in the action, and thus gives rise to a boundary term at $x^a \rightarrow \infty$, which we discard. We obtain
\begin{align}
 S_2 \left[ \tensor[^n]{\hat{\gamma}}{_{ab}}, \n\chi, \n\phi \right] ={}& \sum_{n = 0}^N \int_{\rn} d^5 \lrn x \sqrt{-\hat{\gamma}} \frac{1}{4 \kapfs \lrn{k} \lrn{P}} \frac{\partial}{\partial y} \left\{ e^{\n\chi} R^{(4)} + \frac{3}{2} e^{\n\chi} (\nabla^a \n\chi) (\nabla_a \n\chi) \right\} \nonumber
\\
 &{} + \sum_{n = 0}^{N-1} \n[_m]S \left[ e^{\n\chi (x^a, n)} \tensor{\hat{\gamma}}{_{ab}}, \n\phi \right].
\end{align}
Integrating over $y$, we find boundary terms at each brane and at $y = \pm \infty$.
We note that the integral converges in the first and last regions because of the choices $P_0 = +1$ and $P_N = -1$, and so the terms evaluated at $\pm \infty$ vanish. We can rearrange the remaining terms into a sum over the branes.
\begin{align}
 S_2 \left[ \tensor[^n]{\hat{\gamma}}{_{ab}}, \n\chi, \n\phi \right] ={}& \sum_{n = 0}^{N-1} \int d^4 x \sqrt{-\hat{\gamma}} \frac{1}{4 \kapfs} \left( \frac{1}{\lrn{k} \lrn{P}} - \frac{1}{\lrnp k \lrnp P} \right) \left[ e^{\n\chi} R^{(4)} + \frac{3}{2} e^{\n\chi} (\nabla^a \n\chi) (\nabla_a \n\chi) \right]_{y=n} \nonumber
\\
 & {} + \sum_{n = 0}^{N-1} \n[_m]S \left[ e^{\n\chi (x^a, n)} \tensor{\hat{\gamma}}{_{ab}}, \n\phi \right]. \label{ActionBeforeRedefs}
\end{align}

\subsection{Field Redefinitions\label{FieldRedefs}}
We now have a four-dimensional Ricci scalar and a number of scalar fields, $\n\chi(x^a, n)$, whose values depend on the distance from the first brane to the $(n+1)^{\mathrm{th}}$ brane. We modify our previous notational conventions as follows. We redefine the fields $\n\chi$ and $\hat{\gamma}_{ab}$ via
\begin{align}
\hat{\gamma}_{ab} \rightarrow {}& e^{f(x^a)} \hat{\gamma}_{ab}(x^a) \\
\n\chi(x^a, y) \rightarrow {}& \n\chi(x^a,y) - f(x^a)
\end{align}
where $f(x^a)$ is defined in Eq. \eqref{chidef}. Thus, we no longer impose the constraint that $\det (\hat{\gamma}) = -1$; instead we have $\det (\hat{\gamma}) = - \exp (4 f)$. With the relaxation of this constraint, a Lagrange multiplier term in the action to enforce the constraint is no longer necessary. As $\chi$ will only be evaluated on the branes from now on, we further depart from our previous conventions, and write $\chi_n(x^a)$ to represent $\n\chi(x^a, n)$. Our new conventions also enforce $\chi_0 = 0$, which means that the metric $\hat{\gamma}_{ab}$ is the Jordan frame metric of the first brane, ${\cal B}_0$. Note that the five-dimensional metric is unchanged by these notational changes.

For convenience, we define the quantities
\begin{align}
A_n ={}& \left| \frac{1}{k_n P_n} - \frac{1}{k_{n+1} P_{n+1}} \right|, \label{DefofA}
\\
B_n ={}& \frac{A_n}{A_0}, \label{DefofB}
\\
\epsilon_n ={}& \mathrm{sgn} \left( \frac{1}{k_n P_n} - \frac{1}{k_{n+1} P_{n+1}} \right), \label{epsilondef}
\end{align}
for $0 \le n \le N-1$. It is useful to note that $\epsilon_n$ can be written as
\begin{align}
\epsilon_n 
={}& - \mathrm{sgn} \left( \sigma_n P_n P_{n+1} \right) \label{epsilonuseful}
\end{align}
by using Eq. \eqref{branetunings}. We define an effective four-dimensional gravitational constant $\kappa_4^2$ that is measured by observers on the first brane, ${\cal B}_0$, by
\begin{equation}
\frac{1}{2 \kappa_4^2} = \frac{1}{4 \kappa_5^2} A_0. \label{4dNewtonConst}
\end{equation}
Finally, we define fields $\psi_n$ by
\begin{align}
\psi_n ={}& \sqrt{B_n e^{\chi_n}} \label{psidef}
\end{align}
for $1 \le n \le N-1$. The domain of $\psi_n$ is the positive reals. Rewriting our action in terms of these new variables (and suppressing the subscript `2' from now on), we obtain
\begin{align}
 S \left[\hat{\gamma}_{ab}, \lrn \psi, \n\phi\right] ={}& \int d^4 x \sqrt{-\hat{\gamma}} \frac{\epsilon_0}{2 \kappa_4^2} \left[ R^{(4)} \left( 1 + \sum_{n = 1}^{N-1} \epsilon_0 \epsilon_n \psi_n^2 \right) + 6 \sum_{n = 1}^{N-1} \epsilon_0 \epsilon_n (\nabla^a \psi_n) (\nabla_a \psi_n) \right] \nonumber
\\
 & {} + \tensor[^0]S{_m}[\tensor{\hat{\gamma}}{_{ab}}, \tensor[^0]{\phi}{}] + \sum_{n = 1}^{N-1} \n[_m]S \left[ \frac{\psi_n^2}{B_n} \hat{\gamma}_{ab}, \n\phi \right]. \label{firstjordanaction}
\end{align}
This is the four-dimensional effective action in the Jordan conformal frame of the first brane, ${\cal B}_0$. Note that the target space metric, determined by the kinetic energy term for the scalar fields, is flat, and the target space manifold is a subset of the quadrant of $\mathbb{R}^{N-1}$ in which all the coordinates $\lrn \psi$ are positive, bearing in mind that each $\lrn \psi$ will be bounded either above or below by their definition \eqref{psidef} and Eq. \eqref{chidef}.

\subsection{Transforming to the Einstein Conformal Frame\label{Einsteintransform}}
The Einstein conformal frame metric for the action \eqref{firstjordanaction} is $g_{ab} = \hat{\gamma}_{ab} |\Theta|$, where $\Theta$ is given by
\begin{align}
\Theta = 1 + \sum_{n = 1}^{N-1} \epsilon_0 \epsilon_n \psi_n^2. \label{thetadef}
\end{align}
The four-dimensional effective action becomes
\begin{align}
 S \left[g_{ab}, \lrn \psi, \n\phi\right] ={}& \int d^4 x \sqrt{- g} \, \frac{\epsilon_0 \, \mathrm{sgn}(\Theta)}{2 \kappa_4^2} \left[ \tilde{R}^{(4)}[g] - \frac{3}{2 \Theta^{2}} (\tilde{\nabla}^a \Theta)(\tilde{\nabla}_a \Theta) + 6 \sum_{n = 1}^{N-1} \frac{\epsilon_0 \epsilon_n}{\Theta} (\tilde{\nabla}^a \psi_n) (\tilde{\nabla}_a \psi_n) \right] \nonumber
\\
 & {} + \tensor[^0]S{_m}\left[\frac{1}{|\Theta|}\tensor{g}{_{ab}}, \tensor[^0]{\phi}{}\right] + \sum_{n = 1}^{N-1} \n[_m]S \left[ \frac{\psi_n^2}{B_n | \Theta |} g_{ab}, \n\phi \right], \label{effectiveaction}
\end{align}
where tildes refer to the metric $g_{ab}$. Note that the kinetic energy terms in this action \eqref{effectiveaction} have apparent divergences at $\Theta = 0$. However, for any given set of signs $\epsilon_n$ (which correspond to a choice of model), it can be shown that $|\Theta|$ is bounded away from zero. This arises because of the way each $\lrn\psi$ is bounded either above or below.

%% file: analysis.tex
\section{Analysis of the Action\label{secanalysis}}
In this section, we analyze the four-dimensional effective action \eqref{effectiveaction} in a variety of cases. We begin with the cases of one and two branes, which serve to highlight some features of the model in the general case. In these special cases, our results reduce to previously known results. We then analyze the general situation.

\subsection{One Brane Case}
In the one brane case, the effective action simplifies greatly.
\begin{align}
 S [g_{ab}, \tensor[^0]{\phi}{}] ={}& \int d^4 x \sqrt{- g} \frac{\epsilon_0}{2 \kappa_4^2} \tilde{R}^{(4)}[g] + \tensor[^0]{S}{_m} \left[ g_{ab}, \tensor[^0]{\phi}{} \right].
\end{align}
The four-dimensional effective action is just general relativity ($\epsilon_0 = +1$ if the brane has positive tension). This corresponds to the RS-II model \cite{Randall1999a}.

\subsection{Two Brane Case}
Here the parameter of importance is $\epsilon_0 \epsilon_1$, which from Eqs. \eqref{branetunings}, \eqref{boundaryp} and \eqref{epsilondef} is given by
\begin{align}
\epsilon_0 \epsilon_1 
={}& - \mathrm{sgn} \left( \sigma_0 \sigma_1 \right).
\end{align}
With $P_0$ and $P_2$ predetermined, it is possible for one brane tension to be negative, but not both. Therefore $\epsilon_0 \epsilon_1$ is positive if there is a negative tension brane, and is negative if both branes have positive tension. Using the definition \eqref{thetadef} of $\Theta$, the action \eqref{effectiveaction} becomes
\begin{align}
 S [g_{ab}, \psi_1, \tensor[^0]{\phi}{}, \tensor[^1]{\phi}{}] ={}& \int d^4 x \sqrt{- g} \; \frac{\epsilon_0 \, \mathrm{sgn} (1+\epsilon_0 \epsilon_1 \psi_1^2)}{2 \kappa_4^2} \left[ \tilde{R}^{(4)}[g] + 6 \frac{\epsilon_0 \epsilon_1}{(1 + \epsilon_0 \epsilon_1 \psi_1^2)^2} (\tilde{\nabla}^a \psi_1) (\tilde{\nabla}_a \psi_1) \right] \nonumber
\\
 & {} + \tensor[^0]{S}{_m} \left[ \frac{1}{|1 + \epsilon_0 \epsilon_1 \psi_1^2|} g_{ab}, \tensor[^0]{\phi}{} \right] + \tensor[^1]{S}{_m} \left[ \frac{\psi_1^2}{B_1 |1 + \epsilon_0 \epsilon_1 \psi_1^2|} g_{ab}, \tensor[^1]{\phi}{} \right]. \label{twobraneaction}
\end{align}

\subsubsection{Positive Brane Tensions}
When both branes have positive tension, $\epsilon_0 \epsilon_1 = -1$. Which of $\epsilon_0$ and $\epsilon_1$ is negative depends on the sign of $\Theta$. Combining Eqs. \eqref{thetadef} and \eqref{psidef},
\begin{align}
\Theta = 1 - B_1 e^{\chi_1}. \label{psi1ex}
\end{align}
From Eqs. \eqref{boundaryp} and \eqref{epsilonuseful}, we see that
\begin{align}
\epsilon_0 = - \epsilon_1 = - \mathrm{sgn} (P_1).
\end{align}
Combining this with Eq. \eqref{chidef} and recalling that $\chi_0 = 0$, we see that the exponential function in Eq. \eqref{psi1ex} is greater than unity for $P_1 = +1$, and less than unity for $P_1 = -1$. If $P_1 = +1$, then the brane tensions [Eq. \eqref{branetunings}] require that $k_0 > k_1$, and we see that $B_1 > 1$, giving $\Theta < 0$ for $\epsilon_0 = -1$, $\epsilon_1 = +1$. If $P_1 = -1$, then the brane tensions dictate that $k_0 < k_1$. Thus, in this case, $B_1 < 1$, and so $\Theta > 0$ for $\epsilon_0 = +1$, $\epsilon_1 = -1$.

Assuming that $0 < \psi_1 < 1$ ($\Theta > 0$, $P_1 = -1$, $\epsilon_0 = +1$), we define
\begin{align}
\varphi ={}& \E \tanh^{-1} (\psi_1)
\end{align}
where
\begin{align}
\E ={}& \frac{\sqrt{6}}{\kappa_4}.
\end{align}
The domain of $\varphi$ is $0$ to $\infty$.
The action \eqref{twobraneaction} then becomes
\begin{align}
 S [g_{ab}, \psi_1, \tensor[^0]{\phi}{}, \tensor[^1]{\phi}{}] ={}& \int d^4 x \sqrt{- g} \left[ \frac{1}{2 \kappa_4^2} \tilde{R}^{(4)}[g] - \frac{1}{2} (\tilde{\nabla}^a \varphi) (\tilde{\nabla}_a \varphi) \right] \nonumber
\\
 & {} + \tensor[^0]{S}{_m} \left[ \cosh^2 \left(\frac{\varphi}{\E} \right) g_{ab}, \tensor[^0]{\phi}{} \right] + \tensor[^1]{S}{_m} \left[ \frac{1}{B_1} \sinh^2 \left(\frac{\varphi}{\E} \right) g_{ab}, \tensor[^1]{\phi}{} \right]. \label{PosBranes1}
\end{align}

Requiring that the branes do not intersect or overlap gives
\begin{align}
0 < \psi_1 < \sqrt{B_1} ={}& \sqrt{\frac{1 - k_1/k_2}{1 + k_1/k_0}}. \label{intersectcondition}
\end{align}
Note that $k_1 < k_2$ to satisfy Eq. \eqref{branetunings}, and that $\sqrt{B_1} < 1$ (responsible for $\Theta > 0$). Thus, Eq. \eqref{intersectcondition} is a more stringent constraint than $0 < \psi_1 < 1$.

In the situation where $\psi_1 > 1$ ($\Theta < 0$, $P_1 = +1$, $\epsilon_1 = +1$), we define
\begin{align}
\varphi ={}& \E \tanh^{-1} \left(\frac{1}{\psi_1}\right).
\end{align}
The domain of $\varphi$ is from $0$ to $\infty$.
The action \eqref{twobraneaction} then becomes
\begin{align}
 S [g_{ab}, \psi_1, \tensor[^0]{\phi}{}, \tensor[^1]{\phi}{}] ={}& \int d^4 x \sqrt{- g} \left[ \frac{1}{2 \kappa_4^2} \tilde{R}^{(4)}[g] - \frac{1}{2} (\tilde{\nabla}^a \varphi) (\tilde{\nabla}_a \varphi) \right] \nonumber
\\
 & {} + \tensor[^0]{S}{_m} \left[ \sinh^2 \left(\frac{\varphi}{\E} \right) g_{ab}, \tensor[^0]{\phi}{} \right] + \tensor[^1]{S}{_m} \left[ \frac{1}{B_1} \cosh^2 \left(\frac{\varphi}{\E} \right) g_{ab}, \tensor[^1]{\phi}{} \right], \label{PosBranes2}
\end{align}
which coincides with the previous action \eqref{PosBranes1} if we swap the actions $\tensor[^0]{S}{_m}$ and $\tensor[^1]{S}{_m}$ and rescale units in each matter action by factors of $B_1^{\pm 1/2}$.

The constraint on the radion field we impose to ensure that the branes do not overlap in this case is
\begin{align}
\psi_1 > \sqrt{B_1} ={}& \sqrt{\frac{1 + k_1/k_2}{1 - k_1/k_0}} > 1,
\end{align}
where $k_1 < k_0$ from the brane tunings (Eq. \eqref{branetunings}).

The two cases $P_1 = -1$, described by the action \eqref{PosBranes1}, and $P_1 = +1$, described by the action \eqref{PosBranes2}, lie on opposite sides of $\Theta = 0$. Each set of values $\epsilon_0, \ldots, \epsilon_{N-1}$ gives different theories, and each theory comes with its own field space.

The actions \eqref{PosBranes1} and \eqref{PosBranes2} coincide with formulae in the literature for the action for the RS-I model, up to a rescaling of units \cite{Randall1999, Chiba2000, Goldberger2000} (also, c.f. Eq. \eqref{RSIendaction}). They describe a Brans-Dicke like scalar-tensor theory of gravity, with matter on each brane having a different coupling strength to the scalar component.

\subsubsection{One Negative Brane Tension}
If $\epsilon_0 \epsilon_1 = 1$ then $\Theta > 0$ always, and by requiring the conditions \eqref{boundaryp}, both $\epsilon_0$ and $\epsilon_1$ must be positive. We define
\begin{align}
\varphi ={}& \E \tan^{-1} (\psi_1),
\end{align}
where the domain of $\varphi$ is $0$ to $(\pi/2) \E$.
The action \eqref{twobraneaction} becomes
\begin{align}
 S [g_{ab}, \psi_1, \tensor[^0]{\phi}{}, \tensor[^1]{\phi}{}] ={}& \int d^4 x \sqrt{- g} \left[ \frac{1}{2 \kappa_4^2} \tilde{R}^{(4)}[g] + \frac{1}{2} (\tilde{\nabla}^a \varphi) (\tilde{\nabla}_a \varphi) \right] \nonumber
\\
 & {} + \tensor[^0]{S}{_m} \left[ \cos^2 \left(\frac{\varphi}{\E} \right) g_{ab}, \tensor[^0]{\phi}{} \right] + \tensor[^1]{S}{_m} \left[ \frac{1}{B_1} \sin^2 \left(\frac{\varphi}{\E} \right) g_{ab}, \tensor[^1]{\phi}{} \right].
\end{align}
Note that $\varphi$ is a ghost field, which gives rise to the usual instability associated with a negative tension brane.

\subsection{General Case of N branes\label{GeneralCase}}
In the general case of $N$ branes with $N>2$, we have $N-1$ scalar fields $\psi_1, \ldots, \psi_{N-1}$. It will be useful to regard the $(N-1)$-dimensional space of field configurations as an $N-1$-dimensional hypersurface in an $N$-dimensional space with coordinates $\psi_1, \ldots, \psi_{N}$, where
\begin{align}
\psi_N \equiv \sqrt{-\epsilon_0 \epsilon_N \Theta} = \sqrt{ \left| 1 + \sum_{n = 1}^{N-1} \epsilon_0 \epsilon_n \psi_n^2 \right| } \label{psiNdef}
\end{align}
by Eq. \eqref{thetadef}. We also define $\epsilon_N = - \epsilon_0 \mathrm{sgn}(\Theta)$.
The action \eqref{twobraneaction} becomes
\begin{align}
 S [g_{ab}, \psi_n, \n\phi] ={}& \int d^4 x \sqrt{- g} \; \epsilon_0 \, \mathrm{sgn} (\Theta) \left[ \frac{1}{2 \kappa_4^2} \tilde{R}^{(4)}[g] + \frac{1}{2} \E^2 \sum_{n = 1}^N \frac{\epsilon_0 \epsilon_n}{\Theta} (\tilde{\nabla}^a \psi_n) (\tilde{\nabla}_a \psi_n) \right] \nonumber
 \\
 {}& + \tensor[^0]S{_m}\left[\frac{1}{|\Theta|}\tensor{g}{_{ab}}, \tensor[^0]{\phi}{}\right] + \sum_{n = 1}^{N-1} \n[_m]S \left[ \frac{\psi_n^2}{B_n |\Theta|} g_{ab}, \n\phi \right],
\end{align}
where it is understood that the action is constrained by Eq. \eqref{psiNdef} (this constraint may be enforced by a Lagrange multiplier if desired).

The N-dimensional field space metric ($d\Sigma^2$) is now conformally flat (and indeed, is a form of the metric on $N$-dimensional AdS space with an unusual signature). The physical field space metric is obtained by computing the induced $(N-1)$-dimensional metric on the hypersurface \eqref{psiNdef} from the $N$-dimensional metric. Of the $N-1$ fields $\psi_1, \ldots, \psi_{N-1}$, let $P$ with $0 \le P \le N-1$ be the number of fields with $\epsilon_0 \epsilon_n$ positive, and let $M = N-1-P$ be the number of fields with $\epsilon_0 \epsilon_n$ negative. We now relabel the fields based on which have positive kinetic coefficient, and which have negative kinetic coefficient. Call the fields $p_i$, $1 \le i \le P$, and $m_j$, $1 \le j \le M$. Following the convention in Eq. \eqref{generalEinsteinframe} for the metric on the $N$-dimensional target space of the scalar fields, we can write it as
\begin{align}
\frac{\Theta}{\E^2} d\Sigma^2 ={}& - \sum_{n=1}^{N} \epsilon_0 \epsilon_n d\psi_n^2
\\
={}& - \sum_{i=1}^P dp_i^2 + \sum_{j=1}^M dm_j^2 + \mathrm{sgn}({\Theta}) d\psi_N^2.
\end{align}
As the metric on the positive (negative) coordinates is Euclidean, we can transform them into spherical polar coordinates. Define new coordinates $\zeta, \theta_1, \ldots, \theta_{P-1}$ and $\eta, \lambda_1, \ldots, \lambda_{M-1}$, such that
\begin{subequations}
\label{littlefieldeqns}
\begin{align}
(p_1, \ldots, p_P) = {}& \zeta \left(\cos(\theta_1), \sin(\theta_1) \cos(\theta_2), \ldots, \sin(\theta_1) \sin(\theta_2) \cdots \sin(\theta_{P-1})\right),
\\
(m_1, \ldots, m_N) = {}& \eta \left(\cos(\lambda_1), \sin(\lambda_1) \cos(\lambda_2), \ldots, \sin(\lambda_1) \sin(\lambda_2) \cdots \sin(\lambda_{M-1})\right).
\end{align}
\end{subequations}
In these coordinates, the field space metric is
\begin{equation}
\frac{\Theta}{\E^2} d\Sigma^2 = - d\zeta^2 - \zeta^2 d\Omega_p^2 + d\eta^2 + \eta^2 d\Omega_m^2 + \mathrm{sgn}({\Theta}) d\psi_N^2,
\end{equation}
where $d\Omega_p^2 = d\theta_1^2 + \sin^2(\theta_1) d\theta_2^2 + \ldots$ is the metric on the unit $(P-1)$-sphere, and similarly for $d\Omega_m^2$. All of the angular fields ($\theta_i$ and $\lambda_j$) have a domain of $0$ to $\pi/2$, as each $\psi_n$ is positive. We will refer to this region as the ``positive quadrant'' of the maximally extended fieldspace (where the domains of the angular fields are from $0$ to $\pi$ or $2 \pi$ as per usual). The following relationships hold:
\begin{align}
\zeta^2 ={}& \sum_{i=1}^P p_i^2,
\\
\eta^2 ={}& \sum_{j=1}^M m_j^2,
\\
\Theta ={}& 1 + \zeta^2 - \eta^2. \label{newthetadef}
\end{align}
The domain of $\eta$ and $\zeta$ is from $0$ to $\infty$.

We now compute the physical field space metric on the $(N-1)$-dimensional target space ($d\sigma^2$) using the constraint \eqref{psiNdef}. We obtain
\begin{align}
d \psi_N^2 ={}& \frac{1}{|1 + \zeta^2 - \eta^2|} (\zeta d\zeta - \eta d \eta)^2,
\\
d\sigma^2 
={}& \frac{\E^2}{1 + \zeta^2 - \eta^2} \left[ - d\zeta^2 \left( \frac{1 - \eta^2}{1 + \zeta^2 - \eta^2} \right) - \zeta^2 d\Omega_p^2 + d\eta^2 \left( \frac{1 + \zeta^2}{1 + \zeta^2 - \eta^2} \right) + \eta^2 d\Omega_m^2 - \frac{2 \eta \zeta}{1 + \zeta^2 - \eta^2} d\eta d\zeta \right]. \label{FullFieldMetric}
\end{align}

To summarize, our final result for the four dimensional action is a massless multiscalar-tensor theory in a nonlinear sigma model, of the form
\begin{align}
 S[g_{ab}, \Phi^A, \n\phi] ={}& \int d^4 x \sqrt{-g} \left\{ \frac{1}{2 \kappa_4^2} R [g_{ab}] - \frac{1}{2} \gamma_{AB}(\Phi^C) \nabla_a \Phi^A \nabla_b \Phi^B g^{ab} \right\} \nonumber
\\
 &{} + \sum_{n=0}^{N-1} \n[_m]S \left[ e^{2 \alpha_n (\Phi^C)} g_{ab}, \n\phi \right],
\end{align}
where we assume that $\epsilon_0 \mathrm{sgn}(\Theta) = +1$, which is required for a physical theory, given our conventional choices. The scalar fields $\Phi^A$ are
\begin{align}
\left\{ \Phi^A \right\} = \left\{ \zeta, \theta_1, \ldots, \theta_{P-1}, \eta, \lambda_1, \ldots, \lambda_{M-1} \right\}.
\end{align}
The field space metric $\gamma_{AB}(\Phi^C)$ is given by $d \sigma^2$ [Eq. \eqref{FullFieldMetric}], and the brane coupling functions $\alpha_n (\Phi^C)$ by
\begin{subequations}
\begin{align}
e^{2 \alpha_0} = {}& \frac{1}{|1 + \zeta^2 - \eta^2|},
\\
e^{2 \alpha_n} = {}& \frac{1}{|1 + \zeta^2 - \eta^2|} \frac{\psi_n^2}{B_n}, \ 1 \le n \le N-1,
\end{align}
\end{subequations}
where $B_n$ is given by Eq. \eqref{DefofB}, and $\psi_n$ is defined by the relevant expression in Eq. \eqref{littlefieldeqns}.

%% file: lowenergy.tex
\section{Discussion and Conclusions\label{endsection}}
We have presented a new method to calculate the four-dimensional effective action for five-dimensional models involving $N$ non-intersecting branes in the low energy regime. Although we have only illustrated an application of the method to an uncompactified extra dimension, it is generally applicable, and is expected to work for circularly compactified and orbifolded models also.

\subsection{Domain of Validity of the Four-Dimensional Description\label{HiEnergy}}
We begin our discussion of the domain of validity of the four-dimensional description given by Eq. \eqref{effectiveaction} by recapping the method of computation discussed in Section \ref{SecLengthscale}. Starting from the five-dimensional action $S$, we define a rescaled action $S_\epsilon$ which has the expansion
\begin{equation}
S_\epsilon = \tensor{S}{_0} + \epsilon^2 \tensor{S}{_2}. \label{recapscaledaction}
\end{equation}
In Section \ref{SecLowest} we found the most general solution of $\delta S_0 = 0$, and substituting that solution into $S_2$, gave the four-dimensional action functional of Section \ref{Einsteintransform}\footnote{The action $S_0$ for the solution is zero, assuming the brane tunings \eqref{branetunings}.}.

The basis of our approximation method is the smallness of the bulk radius of curvature $1/k_n$ compared to the radius of curvature ${\cal L}_c$ of the four-dimensional metric $e^{\chi} \hat{\gamma}_{ab}$. However, although this approximation is valid on all the branes, it inevitably breaks down as $y \rightarrow \pm \infty$, far from the branes, as ${\cal L}_c \rightarrow 0$, as discussed in Section \ref{secdiscuss}. It is worth noting that in the special case where all of the induced metrics on the branes are flat and there are no matter fields, the metric ansatz (with $\Phi = \mathrm{const}$) is an exact solution to the five-dimensional Einstein equations, and this breakdown does not occur.

One might expect contributions from the regime far from the branes to invalidate our four-dimensional effective description. However, we expect that the contribution to the action far from the brane will negligibly change the calculation, as in the region in which we expect large departures from the derived metric, the warp factor exponentially suppresses any contributions.

It is possible for our two-lengthscale expansion to break down not only asymptotically, but also in between branes. A number of models (eg, \cite{Carena2005, Flanagan2001a, Damour2002, Karch2001} to cite but a few) discuss ``bounce'' behavior in the warp factor, where it decreases and increases again in between branes, as with a $\cosh^2$ dependence. Typically, this behavior appears when the metric $\hat{\gamma}$ is a curved FRW metric. It is a limitation of our method that this bounce is not evident in our solutions, as it explicitly requires coupling between the $O(1)$ and $O(\epsilon^2)$ components (in particular, the four-dimensional Ricci scalar). Thus, this behavior is excluded by the underlying assumptions of our method, as near the turning point of these bounces, the separation of lengthscales has broken down. We note, however, that $\cosh^2$ behavior is likely to be forbidden in the first or last ($y \rightarrow \pm \infty$) regions by global hyperbolicity. It is also possible to produce $\sinh^2$ behavior in the warp factor. In between branes, this can lead to topologically disconnected regions of spacetime as discussed in \cite{Flanagan2001a}, which we have excluded by assumption. In the first or last regions, correctly accounting for this behavior requires that the integration over the fifth dimension be truncated. However, the contributions to our effective action from integrating beyond these regions is again exponentially suppressed and negligible. In the regime in which the separation of lengthscales is valid, our solutions are in agreement with models displaying these types of behavior.

For black holes, the solution given by our effective action is subject to the Gregory-Laflamme instability \cite{Gregory1993} and the final outcome is uncertain (see \cite{Chamblin2000} and citations thereof). The five-dimensional stability of solutions for which the induced metric on the branes is not nearly flat (eg, neutron stars) is an interesting open question. We conjecture that all the solutions without horizons are stable and are reasonably described by our four-dimensional effective action.

We may also consider the regime in which ${\cal L}_c \ll {\cal L}$, such as will occur a long way away from the branes. In this limit, the physical description would change from being that of decoupled fibres to that of decoupled four-dimensional hypersurfaces [one should solve the $O(\epsilon^2)$ contribution to the action first, and substitute that into the $O(1)$ contribution to the action]. This approach may yield a matched asymptotic expansion approach to obtaining a solution far from the branes. Our method may therefore be useful for investigating the regime between Minkowski space on a brane and a black hole on a brane.

It is important to note that our method does not yield the leading order five-dimensional metric. This can be seen from the fact that our four dimensional action depends only on the fields $\n\chi$ evaluated on the branes, and the values of these fields between the branes are not determined. However, knowledge of the leading order five-dimensional metric is, rather surprisingly, \emph{not} a prerequisite for correctly capturing the leading order four-dimensional dynamics. Most other methods rely on knowledge of the five-dimensional behavior of the metric to calculate the effective four-dimensional equations of motion, and our method is somewhat unique in this regard.

Our method of computation correctly captures the leading order dynamics of the system. However, there will be higher order corrections, suppressed by powers of $\epsilon^2$. In particular, the fields $\n\chi$ and $\n\Phi$ can be expanded as
\begin{subequations}
\label{fieldexpansions}
\begin{align}
\n\chi = {}& \n[^{(0)}]{\chi} + \epsilon^2 \n[^{(2)}]{\chi} + O(\epsilon^4),
\\
\n\Phi = {}& \n[^{(0)}]{\Phi} + \epsilon^2 \n[^{(2)}]{\Phi} + O(\epsilon^4).
\end{align}
\end{subequations}
Throughout this paper, we have dealt only with the fields $\n[^{(0)}]{\chi}$ and $\n[^{(0)}]{\Phi}$. The necessity for the higher order terms can be seen from the exact, five-dimensional equations of motion, which are derived in Appendix \ref{AppExactEOMS}. For example, the exact Israel junction conditions are given by Eq. \eqref{branetuningsepsilon}. If we substitute the expansions \eqref{fieldexpansions} into Eq. \eqref{branetuningsepsilon}, and use \eqref{jumpconditions} [with \n\chi\ and \n\Phi\ replaced by $\n[^{(0)}]{\chi}$ and $\n[^{(0)}]{\Phi}$] together with the brane tuning conditions \eqref{branetunings}, we find that the higher order corrections $\n[^{(2)}]{\chi}$ and $\n[^{(2)}]{\Phi}$ are related to the matter stress energy tensors on the brane. Our results confirm the suggestion of Kanno and Soda that these higher order corrections do not affect the four-dimensional effective action to leading order \cite{Kanno2005a}.

\subsection{Models Which Violate the Brane Tension Tunings\label{BraneTensions}}
If a brane's tension is adjusted so as to violate the tuning condition \eqref{branetunings}, then it is possible to view the situation as having either detuned brane tensions or detuned bulk cosmological constants. For accounting purposes, it is simpler to think of the bulk cosmological constants as being detuned. When this occurs, the exact equations of motion in the bulk \eqref{exacteom1} imply that a nonzero Ricci curvature is induced to compensate for the detuning. Exact solutions have been calculated in highly symmetric cases, see for example Ref. \cite{Karch2001}. In general, the exact nature of the perceived detuning is nontrivial, as the bulk cosmological constants on either side of the offending brane(s) can appear detuned by different amounts to compensate.

If the deviation from the brane tuning conditions is small [$\Delta \sigma / \sigma^T = O(\epsilon^2)$], then we can approximate the contribution to the four-dimensional effective action as
\begin{align}
\Delta S = - \sum_{n=0}^{N-1} \int d^4 x \sqrt{-\n{h}} (\sigma_n - \sigma_n^T),
\end{align}
where $\sigma_n^T$ is the tuned value for the $n^{\mathrm{th}}$ brane, given by \eqref{branetunings}. This approximation is of the same order as the other approximations we have made in our method. The net result is then an effective cosmological constant on each brane, given by
\begin{align}
\Lambda^{(4)}_n = \sigma_n - \sigma_n^T,
\end{align}
which vanishes when the brane tensions are tuned. [Note: This differs from the result given in the literature for the RS-II model, see for example Ref. \cite{Maartens2004}, but the difference is $O(\epsilon^4)$].

If the detuning of a brane's tension from its tuned value should become too large [$O(1)$ rather than $O(\epsilon^2)$], then the curvature induced by the four-dimensional effective cosmological constant can cause the radius of curvature on a slice of constant $y$ close to the branes to violate the approximations used in our method, which implies that our four-dimensional effective action will not be a good description of a system in this regime.


\subsection{Multigravity\label{Discussions}}
Theories with more than one independent dynamical tensor field are called multigravity theories; see the general discussion in Damour and Kogan \cite{Damour2002}. The models in this paper may exhibit two forms of multigravity, although we have ignored one of them entirely.

The first form of multigravity is the possible existence of a second tensor field, given by the matrix $\mathbf{B}(x^a)$ in Eq. \eqref{gammasoln}. We argued in Section \ref{SecClasses} that this form of multigravity is likely forbidden.

The second form of multigravity arises from the the fact that outside of the low energy regime, the models will contain Kaluza-Klein graviton modes. These modes will have masses that are formally of order ${\cal L}^{-1}$, but may be much lighter due to exponential suppression factors, and so be phenomenologically important (so-called ``ultra-light modes'')\cite{Kogan2000, Kogan2001}. Our method of analysis automatically excludes all massive fields (formally, we take $\epsilon$ sufficiently small to overcome any large exponential factors), so we have neglected all graviton Kaluza-Klein modes. It is likely that some of these modes are in fact ultralight in our model, as in the analyses of Damour and Kogan \cite{Kogan2000, Kogan2001, Damour2002}.

\subsection{Conclusions}
The method we have developed is a useful tool for investigating the four-dimensional, low energy behavior of five-dimensional braneworld models. We have illustrated the method for a system of $N$ branes in an uncompactified extra dimension. We intend to apply the method developed here to orbifolded models. We also plan to investigate experimental constraints on the model illustrated here, especially solar system constraints. Finally, we will investigate dark matter models obtained by placing matter fields on various branes.

\begin{acknowledgements}
We would like to thank S.-H. Henry Tye and Ira Wasserman for helpful discussions. This research was supported in part by NSF grants 0757735 and 0555216, and NASA grant NNX08AH27G.
\end{acknowledgements}

%% file: exact.tex
\section{Five-Dimensional Ricci Scalars and Exact Equations of Motion\label{AppExactEOMS}}
Here we present the dimensionally reduced Ricci scalar and the exact equations of motion for the action \eqref{CompleteAction}. We include the order at which terms appear in terms of our scaling parameter, $\epsilon$.

\subsection{Dimensional Reduction of the Ricci Scalar}
The constraint $\det \hat{\gamma} = -1$ may be enforced either at the level of the equations of motion, or by using a Lagrange multiplier.

If the constraint $\det \hat{\gamma} = -1$ is being enforced at the level of the equations of motion, then it is simplest to compute the equations of motion using the metric \eqref{firstmetric}, and then perform a conformal transformation on the quantities in the equations of motion. In this metric, the five-dimensional Ricci scalar is given by
\begin{align}
\n[^{(5)}]R ={}& \epsilon^2 \left(\n[^{(4)}]R - \frac{2 \nabla^a \nabla_a \n\Phi}{\n\Phi}\right) - \frac{\n[^{ab}]\gamma \n[_{ab,yy}]\gamma}{\n[^2]\Phi} + \n[^{ab}]\gamma \n[_{ab,y}]\gamma \frac{\n[_{,y}]\Phi}{\n[^3]\Phi} \nonumber
\\
{}& - \frac{1}{4 \n[^2]\Phi} \left( \n[^{ab}]\gamma \n[_{ab,y}]\gamma \right)^2 + \frac{3}{4 \n[^2]\Phi} \n[^{ab}]\gamma \n[_{ac,y}]\gamma \n[^{cd}]\gamma \n[_{db,y}]\gamma , \label{Ricci1expansion}
\end{align}
where covariant derivatives and the four-dimensional Ricci scalar are those associated with $\n[_{ab}]\gamma$.

For the constraint $\det \hat{\gamma} = -1$ to be enforced at the level of the action, a Lagrange multiplier term must be added to the action
\begin{align}
\Delta S = \sum_{n=0}^{N} \int_\rn d^5 x_n \n\lambda (x^a, y) \left(\sqrt{-\n{\hat{\gamma}}} - 1\right), \label{Lagrangemults}
\end{align}
where $\n\lambda (x^a, y)$ are the Lagrange multiplier fields. Using the metric \eqref{scaledmetric}, the five-dimensional Ricci scalar is given by
\begin{align}
\n[^{(5)}]R ={}& \epsilon^2 e^{-\n\chi} \left(\n[^{(4)}]R - 3 \nabla^a \nabla_a \n\chi  - \frac{3}{2} (\nabla^a \n\chi) (\nabla_a \n\chi) - \frac{2 \nabla^a \nabla_a \n\Phi}{\n\Phi} - \frac{2 (\nabla^a \n\chi) (\nabla_a \n\Phi)}{\n\Phi} \right) \nonumber
\\
{}& + \frac{1}{\n\Phi^2} \left( -\frac{1}{4}\n[^{ab}]{\hat{\gamma}} \n[_{ac,y}]{\hat{\gamma}} \n[^{cd}]{\hat{\gamma}} \n[_{db,y}]{\hat{\gamma}} - 5 (\n[_{,y}]\chi)^2 - 4 \n[_{,yy}]\chi + 4 \frac{\n[_{,y}]\Phi}{\n\Phi} \n[_{,y}]\chi \right) , \label{Ricci2expansion}
\end{align}
where covariant derivatives and the four-dimensional Ricci scalar are those associated with $\n[_{ab}]{\hat{\gamma}}$. To obtain this form, we use the following two formulae which may be derived from the fact that $\det (\n[_{ab}]{\hat{\gamma}}) = -1$:
\begin{align}
\n[^{ab}]{\hat{\gamma}} \; \n[_{ab, y}]{\hat{\gamma}} ={}& 0, \label{gammamotion}
\\
\n[^{ab}]{\hat{\gamma}} \; \n[_{ab, yy}]{\hat{\gamma}} ={}& \n[^{ab}]{\hat{\gamma}} \; \n[_{bc, y}]{\hat{\gamma}} \n[^{cd}]{\hat{\gamma}} \; \n[_{da, y}]{\hat{\gamma}}. \label{gammaswaps}
\end{align}
The complete action (with $\epsilon$ scaling and Lagrange multipliers) is given by Eq. \eqref{scaledepsilonaction}.

\subsection{Varying the Action}
We use $\n[_{ab}]{\hat{\gamma}}$ to compute covariant derivatives, the four-dimensional Ricci scalar ($\n[^{(4)}]{R}$) and the four-dimensional Einstein tensor ($\n[^{(4)}_{ab}]{G}$). Indices will also be raised and lowered using this metric.

Varying the action \eqref{CompleteAction} with respect to $\n\Phi$, we find the bulk equation of motion
\begin{align}
\epsilon^2 e^{-\n\chi} \left( \n[^{(4)}]{R} - \frac{3}{2} ({\nabla}^a \n\chi) ({\nabla}_a \n\chi) - 3 {\nabla}^a{\nabla}_a \n\chi \right) - \frac{3}{\n[^2]{\Phi}} \n[^2_{,y}]\chi + \frac{1}{4 \n[^2]{\Phi}} \n[^{ab}]{\hat{\gamma}}\n[_{bc,y}]{\hat{\gamma}}\n[^{cd}]{\hat{\gamma}}\n[_{da,y}]{\hat{\gamma}} - 2 \kapfs \lrn \Lambda = 0. \label{exacteom1}
\end{align}
From combining the variations with respect to $\n[_{ab}]{\hat{\gamma}}$ and $\n\chi$ (after eliminating the Lagrange multiplier by tracing over the $\n[_{ab}]{\hat{\gamma}}$ equation of motion, or enforcing $\det \hat{\gamma} = -1$ on the equations of motion), we obtain a traceless tensor equation of motion in the bulk
\begin{align}
{}& \frac{1}{2} \n[^2]{\Phi} \epsilon^2 e^{- \n\chi} \left( 4 \n[^{(4)}_{ab}]{G} + \n[_{ab}]{\hat{\gamma}} \n[^{(4)}]{R} + 2 ({\nabla}_a \n\chi) ({\nabla}_b \n\chi) - \frac{1}{2} \n[_{ab}]{\hat{\gamma}} ({\nabla}^c \n\chi) ({\nabla}_c \n\chi) - 4 {\nabla}_a{\nabla}_b \n\chi + \n[_{ab}]{\hat{\gamma}} {\nabla}^c{\nabla}_c \n\chi \right) \nonumber
\\
{}& + \frac{3}{2} \n{\Phi} \epsilon^2 e^{- \n\chi} \left( - 4 {\nabla}_a {\nabla}_b \n{\Phi} + 4 ({\nabla}_{(a} \n\Phi) ({\nabla}_{b)} \n\chi) + \n[_{ab}]{\hat{\gamma}} {\nabla}^c {\nabla}_c \n{\Phi} - \n[_{ab}]{\hat{\gamma}} ({\nabla}_c \n{\Phi}) ({\nabla}^c \n{\chi}) \right) \nonumber
\\
{}& - \n[_{ab,yy}]{\hat{\gamma}} + \frac{\n[_{,y}]{\Phi}}{\n{\Phi}} \n[_{ab,y}]{\hat{\gamma}} - 2 \n[_{,y}]{\chi} \n[_{ab,y}]{\hat{\gamma}} + \n[_{ac,y}]{\hat{\gamma}}\n[^{cd}]{\hat{\gamma}}\n[_{db,y}]{\hat{\gamma}} = 0,
\end{align}
and a scalar equation of motion in the bulk
\begin{align}
{}& \frac{1}{2} \n[^2]{\Phi} \epsilon^2 e^{-\n\chi} \left( - \n[^{(4)}]{R}  + \frac{3}{2} ({\nabla}^c \n\chi) ({\nabla}_c \n\chi) + 3 {\nabla}^c{\nabla}_c \n\chi + \frac{5}{\n\Phi} {\nabla}^c {\nabla}_c \n{\Phi} + \frac{5}{\n\Phi} ({\nabla}_c \n\Phi) ({\nabla}^c \n\chi) \right) \nonumber
\\
{}& + \frac{1}{4} \n[^{ab}]{\hat{\gamma}}\n[_{ab,yy}]{\hat{\gamma}} + 3 \n[_{,yy}]{\chi} + 3 (\n[_{,y}]{\chi})^2 - 3 \frac{\n[_{,y}]{\Phi}}{\n{\Phi}} \n[_{,y}]{\chi} + 2 \n[^2]{\Phi} \kapfs \lrn \Lambda = 0.
\end{align}
These variations also give rise to the boundary conditions on the branes
\begin{align}
\frac{1}{\n\Phi} \n[_{ab, y}]{\hat{\gamma}} - \frac{1}{\np\Phi} \np[_{ab, y}]{\hat{\gamma}} = 2 \kapfs \epsilon^2 e^{-\n\chi} \left( \n[_{ab}]{T} - \n[_{ab}]{\hat{\gamma}} \frac{1}{4} \, \n[^{cd}]{\hat{\gamma}} \n[_{cd}]{T} \right),
\end{align}
and
\begin{align}
- \frac{3 \n[_{,y}]{\chi}}{\n\Phi} + \frac{3 \np[_{,y}]{\chi}}{\np\Phi} + 2 \kapfs \lrn{\sigma} = \frac{1}{2} \kapfs \epsilon^2 e^{-\n\chi} \, \n[^{ab}]{\hat{\gamma}} \n[_{ab}]{T}. \label{branetuningsepsilon}
\end{align}
Here, the four-dimensional stress energy tensors on the branes ($\n[_{ab}]{T}$) are defined by
\begin{align}
\n[_m]{S} [\n[_{ab}]{h} + \delta \n[_{ab}]{h}, \n\phi] = \n[_m]{S} [\n[_{ab}]{h}, \n\phi]
- \frac{1}{2} \int d^4 w_n \sqrt{-\n{h}} \n[_{ab}]{T} \delta \n[^{ab}]{h}.
\end{align}

Note that every factor of $\epsilon^2$ is accompanied by a factor of $\exp({-\n\chi})$. Also note that the $O(1)$ terms in these equations are exactly our equations of motion \eqref{eqmotion2} to \eqref{jumpconditions}.